\documentclass[fleqn,usenatbib]{mnras}

\usepackage{newtxtext,newtxmath}
\usepackage[T1]{fontenc}

\DeclareRobustCommand{\VAN}[3]{#2}
\let\VANthebibliography\thebibliography
\def\thebibliography{\DeclareRobustCommand{\VAN}[3]{##3}\VANthebibliography}

\usepackage{graphicx}	% Including figure files
\usepackage{amsmath}	% Advanced maths commands
\newcommand{\Mvir}{M_{\mathrm{vir}}}
\newcommand{\rvir}{r_{\mathrm{vir}}}
\newcommand{\rs}{r_{\mathrm{s}}}
\newcommand{\tcr}{T_{\mathrm{cr}}}
\newcommand{\rhoc}{\rho_{\mathrm{c}}}
\newcommand{\rh}{r_{\mathrm{h}}}
\newcommand{\rc}{r_{\mathrm{c}}}
\newcommand{\rt}{r_{\mathrm{t}}}
\newcommand{\vesc}{v_{\mathrm{esc}}}
\newcommand{\vescsc}{v_{\mathrm{esc,sc}}}
\newcommand{\vescnfw}{v_{\mathrm{esc,halo}}}
\newcommand{\Msc}{M_{\mathrm{sc}}}
\newcommand{\Og}{\Omega_{\mathrm{\Gamma}}}
\newcommand{\Ol}{\Omega_{\mathrm{\Lambda}}}
\newcommand{\Om}{\Omega_{\mathrm{M}}}
\newcommand{\mmin}{m_{\mathrm{imf,min}}}
\newcommand{\mfmax}{m_{\mathrm{f,max}}}
\newcommand{\mf}{m_{\mathrm{f}}}
\newcommand{\nmerge}{n_{\mathrm{merge}}}
\newcommand{\nrl}{n_{\mathrm{10}}}
\newcommand{\mrl}{m_{\mathrm{10}}}
\newcommand{\rhorl}{\rho_{\mathrm{10}}}
\newcommand{\trxl}{t_{\mathrm{rx10}} }
\newcommand{\fierl}{f_{\mathrm{ie}}}
\newcommand{\fcrl}{f_{\mathrm{coll}}}
\newcommand{\rl}{r_{\mathrm{10}}}
\newcommand{\sigmarl}{\sigma_{\mathrm{10}}}
\newcommand{\rcoll}{r_{\mathrm{coll}}}
\newcommand{\tms}{t_{\mathrm{ms}}}

\newcommand{\tf}{t_{\mathrm{f}}}
\newcommand{\mstar}{m_{\mathrm{star}}}
\newcommand{\mimbh}{m_{\mathrm{imbh}}}
\newcommand{\mbh}{m_{\mathrm{bh}}}
\newcommand{\mvms}{m_{\mathrm{vms}}}
\newcommand{\fpeak}{f_{\mathrm{peak}}}
\newcommand{\adot}{\dot{a}}
\newcommand{\edot}{\dot{e}}

\newcommand{\hcn}{h_{\mathrm{c},n}}
\newcommand{\fnz}{f_{n,z}}
\newcommand{\forb}{f_{\mathrm{orb}}}
\newcommand{\cc}{C}
\newcommand{\vc}{c}

\title[GW of IMBHs in Pop3 clusters]{Gravitational wave of intermediate-mass black holes in Population III star clusters}

\author[Long Wang et al.]{Long Wang,$^{1,2,3}$\thanks{E-mail: wanglong8@mail.sysu.edu.cn (Sysu)}
Ataru Tanikawa,$^{4}$
Michiko Fujii$^{3}$
\\
$^{1}$School of Physics and Astronomy, Sun Yat-sen University, Daxue Road, Zhuhai, 519082, China\\
$^{2}$CSST Science Center for the Guangdong-Hong Kong-Macau Greater Bay Area, Zhuhai, 519082, China \\
$^{3}$Department of Astronomy, Graduate School of Science, The University of Tokyo, 7-3-1 Hongo, Bunkyo-ku, Tokyo 113-0033, Japan\\
$^{4}$Department of Earth Science and Astronomy, College of Arts ans Sciences, The University of Tokyo, 3-8-1 Komaba, Meguro-ku, Tokyo 153-8902, Japan \\
}

\date{Accepted XXX. Received YYY; in original form ZZZ}

\pubyear{2022}

\begin{document}
\label{firstpage}
\pagerange{\pageref{firstpage}--\pageref{lastpage}}
\maketitle

% Abstract of the paper
\begin{abstract}
Previous theoretical studies suggest that the Population III (Pop3) stars tend to form in extremely metal poor gas clouds with approximately $10^5 M_\odot$ embedded in mini dark matter halos.
Very massive stars can form via multiple collisions in Pop3 star clusters and eventually evolve to intermediate-mass black holes (IMBHs).
In this work, we conduct star-by-star $N$-body simulations for modelling the long-term evolution of Pop3 star clusters.
We find that if the mini dark matter halos can survive today, these star clusters can avoid tidal disruption by the galactic environment and can efficiently produce IMBH-BH mergers among a wide range of redshift from 0 to 20.
The average gravitational wave event rate is estimated to be $0.1-0.8~\mathrm{yr}^{-1} \mathrm{Gpc}^{-3}$, and 
approximately $40-80$ percent of the mergers occur at high redshift ($z>6$). 
The characteristic strain shows that a part of low-redshift mergers can be detected by LISA, TianQin, and Taiji, whereas most mergers can be covered by DECIGO and advanced LIGO/VIRGO/Kagra. 
Mergers with pair-instability BHs have a rate of approximately  $0.01-0.15$~yr$^{-1}$~Gpc$^{-3}$, which can explain the GW190521-like events.
%Therefore, GW events in Pop3 star clusters 
%If the mergers occur within 1~Gpc, most of the low-eccentric mergers can be detected by the future space-borne gravitational wave (GW) detectors like LISA, TianQin and Taiji. 
%Thus, GW events originating from Pop3 III star clusters will be important sources for the detected using the future multiband GW detectors. 
%We concludes that Population III star clusters is a suitable environment to produce the IMBH-BH and BBH GW events for the future multiband GW detections.
%The IMBH-BH and BBH GW events originated from Population III clusters will be observed by the future multiband GW detections.
%Star clusters formed in such an environment only contain $10^3-10^4$ stars. 
%Young star clusters like them cannot survive long under the galactic tidal field.
%However, in this work by using \textsc{petar} to perform a large group of star-by-star $N$-body simulations, we find that the mini dark matter halo provides a strong gravitational potential that can keep the clusters surviving until today.
%Meanwhile, if the initial central density of the cluster is high, the hierarchical collisions of massive stars can lead to the formation of intermediate-mass black hole (IMBH), which can interact with stellar mass black hole (BH) and result in IMBH-BH during the long-term evolution.
%We study in detail how the formation of IMBH and the IMBH-BH mergers depend on the initial condition of population III star clusters.
\end{abstract}

% Select between one and six entries from the list of approved keywords.
% Don't make up new ones.

%%%%%%%%%%%%%%%%%%%%%%%%%%%%%%%%%%%%%%%%%%%%%%%%%%

%%%%%%%%%%%%%%%%% BODY OF PAPER %%%%%%%%%%%%%%%%%%

\section{Introduction}

The formation of super massive black holes (SMBHs) is one key question for understanding the cosmological structure formation.
If the massive galaxies grow up via assembly of low-mass ones, the masses of their central SMBHs is also expected to increase via the mergers of binary SMBHs.
The $M-\sigma$ and $M_{\mathrm{BH}}-M_{\mathrm{bulge}}$ relations suggest such co-evolution 
\citep[see][and references therein]{Kormendy2013}.
The high-precision observation of stellar orbits near the Galactic center suggests an SMBH of $4.28\times10^6~M_\odot$  \citep{Ghez1998,Ghez2005,Ghez2008,Gillessen2009,Gillessen2017}.
Recently, the Event Horizon Telescope detect the emission surrounding a SMBH of $6.5\times10^9~M_\odot$ at the center of the elliptical galaxy M87 \citep{EHT2019}.

The Ligo-Virgo-KAGRA (LVK) ground-based gravitational wave (GW) detectors have detected a group of stellar-mass black hole (BH) mergers with a chirp mass of up to approximately $100~M_\odot$ \citep{Abbott2019,Abbott2021,LVK2021}.
The gap between stellar-mass BHs and SMBHs, the intermediate-mass black holes (IMBHs) or the potential seed of SMBHs, with a mass range between $10^3 -10^5~M_\odot$ are still missing conclusive evidence. 
A recent review of IMBH can be found in \cite{Greene2020}.

Considerable efforts have been driven to find the evidence of IMBHs in the extra-galactic environment, including ultraluminous X-Ray sources (ULXs), such as M82 \citep[][]{Kaaret2001,Matsumoto2001,Hopman2004}, spiral galaxy NGC 1313 \citep[][]{Miller2003}, NGC 4395 \citep{Shih2003}, NGC 4559 \citep{Cropper2004}, POX 52 \citep{Barth2004}, NGC 3628 \citep{Strickland2001}, M101 \citep{Kong2004}, Galaxy ESO 243-49 HLX-1 \citep{Farrell2009,Wiersema2010,Davis2011,Lasota2011,Farrell2012,Webb2012}, NGC 5408 X-1 \citep{Strohmayer2009}, NGC 2276 \citep{Mezcua2015} and many other candidates \citep[e.g.,][]{Sutton2012}.
IMBHs can also exist in active galactic nuclei (AGN) \citep[e.g.,][]{Greene2004,Kamizasa2012,Chilingarian2018,Mezcua2018}.
%width-luminosity mass scaling relation candidates in  AGN \citep{Greene2004}, X-ray variability candidates in AGN \citep{Kamizasa2012}, IMBHs candidates by X-ray from accreting disks in AGN \citep{Chilingarian2018}, \citep{Mezcua2018}.
However, the mass estimation of SMBHs or IMBHs in the extra galactic environment  has a large uncertainty. 
It depends on which disk model is assumed \cite[e.g.,][]{Vierdayanti2006}.
\cite{Kording2002} suggested that if X-ray sources do not have isotropic emission, but are relativistically beamed, the required BH masses are reduced to stellar mass.

IMBHs might also exist in dense globular clusters (GCs).
However, it is also challenging to confidently confirm the existence of IMBHs there.
One mannel to detect IMBHs is by measuring the rise of line-of-sight velocity dispersion of central stars, such as  Hubble Space Telescope (HST) observations of M15 \citep[][]{vdMarel2002,Gerssen2002,Gerssen2003} and the HST and Keck observations of G1 \citep{Gebhardt2002,Gebhardt2005}.
Nevertheless, it is difficult to distinguish whether it is a IMBH, a group of stellar-mass BHs and NSs, or just a stellar-mass binary BH (BBH) \citep[e.g.,][]{Hurley2007}. 
If BBHs are detected in GCs, the probability of existing an IMBH is small \citep{Leigh2014}.
\cite{McNamara2003} analysed the proper motion of HST and performed $N$-body models. 
They concluded that there is a weak evidence of IMBH in M15.
\cite{Baumgardt2003a} and \cite{Baumgardt2003b} showed that the $N$-body models of M15 and G1 are consistent with no IMBH.
\cite{Pooley2006} indicated that X-rays from G1 cannot distinguish whether it is an IMBH or a low-mass X-ray binary.

A massive GC such as $\omega$-Centauri is one potential candidate to hold an IMBH \citep{Baumgardt2017}.
The HST and Gemini GMOS-IFU observations found an increase of central velocity dispersion \citep{Noyola2008}.
However, further studies have contradictory results \citep{Anderson2010a,Anderson2010b}.
\cite{Zocchi2017} argued that the radial anisotropy can also explain the velocity dispersion profile instead of an IMBH. 
A following study by using the dynamical model from \cite{Baumgardt2019} also supported this result.

Another series of IMBH detections from the Very Large Telescope (VLT) and HST integrated spectroscopic data show a rise of central velocity dispersion in NGC 6388 \citep{Luetzgendorf2011}, NGC 2808 \citep{Luetzgendorf2012}, NGC 5286 \citep{Feldmeier2013} and other GCs \citep{Luetzgendorf2013}.
However, this evidence is inconclusive. 
The velocity measurement from individual stars based on the HST and the Wide Field Instrument (WFI) shows no rise of the velocity dispersion at the center of NGC 6388 \citep{Lanzoni2013}.

Using pulsar data, \cite{Kzltan2017} claimed that an IMBH exists in 47~Tuc GC, and \cite{Perera2017} suggest an IMBH may exist in NGC 6624, depending on orbital eccentricity. 
\cite{Mann2019} argued that the multi-mass velocity dispersion shows that stellar mass BHs are sufficient to explain the observational data in 47~Tuc GC.
The dynamical model from \cite{Baumgardt2019} suggests no evidence of IMBH in NGC 6624.
No radio signal of IMBHs has been detected in NGC 2808 \citep{Maccarone2008} and other GCs thus far \citep[the MAVERIC project from][]{Tremou2018}.

\cite{Lin2018} discovered a tidal disruption event in an off-center star cluster of a large lenticular galaxy, where an IMBH might exist with the mass ranging from $5\times10^4$ to $10^5 M_\odot$.
Efforts have also been driven to detect IMBH in molecular cloud.
The millimetre-wave emission indicates that a massive IMBH might exist in CO-0.40-0.22 \citep{Oka2017}.
Several other works are not fully covered here.

Future observations, particularly the GW detection, will accurately measure the masses of BHs and probably find the first strong proof of IMBHs. 
%\cite{Cann2018} suggest that the infrared coronal lines from James Webb Space Telescope (JWST) can be used to detect IMBHs in AGNs. 
Space-borne GW detectors, such as LISA \citep{Will2004,LISA2022}, TianQin \citep{Luo2016,Fan2020,Liu2020}, Taiji \citep{Ruan2020}, and DECIGO \citep{Kawamura2011}, can detect the signals of IMBH from a wide region of redshift ($z$), particularly the seed of SMBH with $z>10$.
The GW lensing of IMBHs can also be detected \citep{Lai2018}.
The highly eccentric IMBH-BH merger from Lidov-Kozai (LK) oscillations can also be detected by the ground-based detector like LVK \citep{Fragione2019}.
Combining the space-borne GW detectors and ground-base detectors including LVK, Einstein Telescope \citep{Punturo2010}, and Cosmic Explorer \citep{Reitze2019}, the future multiband detectors can provide rich information of GW from IMBHs \citep{Jani2020}, 

%\cite{Hopman2004,PZ2004} suggests that ULX can be generated by an Roche lobe overflow between an IMBH and a tidal captured star.
%Tidal disruption event by an IMBH in X-ray source NGC 1399 \citep{Irwin2010}.

%Radio emission can be used to identify IMBH from x-ray source \citep{Maccarone2004}.

%An massive IMBH ($>10^5 M_\odot$) may exist in the NSC at NGC 404 \cite{Seth2010}.

%IMBHs may exist in Dwarf Starburst Galaxies Up to Redshift=1.5 \citep{Mezcua2016}.

%Review from \cite{Greene2020}.
Moreover, the formation process of IMBHs should be understood, which is also reflected on the properties of GW signals.
IMBHs can form via a continual merger of stellar-mass objects in dense stellar systems.
In post core-collapse massive star clusters where the central number density $>10^5$ pc$^{-1}$, the continual mergers of stellar-mass BHs can generate IMBH with $M>10^3 M_\odot$ \citep[e.g.,][]{Miller2002,Gultekin2004,Giersz2015,Rizzuto2021,Mapelli2021,Mapelli2022}.
With a higher density, the runaway mergers can rapidly generate IMBH \citep[e.g.,][]{Mouri2002,Giersz2015}.
Massive IMBHs and the seeds of super massive black holes (SMBHs) can form in the circumnuclear giant HII region \citep[e.g.,][]{Taniguchi2000} and in nuclear star clusters \citep[e.g.,][]{Antonini2019,Fragione2020,Kroupa2020,Fragione2022,Rose2022}, where a strong gravitational potential from the galaxy can prevent the escape of compact remnants due to natal kick after supernovae, dynamical ejection after few-body interactions or dynamical recoil kicks due to asymmetric GW.

In an AGN disk around SMBH, IMBH can form via mergers of stars \citep{McKernan2012,McKernan2014}.
In addition, IMBHs and SMBHs may also form via the direct collapse of metal-poor giant gas where the gas cloud undergoes gravitational collapse without the formation of stars at the galactic center \citep[e.g.,][]{Mayer2010,Mayer2015}. 

In a starburst star-forming region with a high central density, the continual mergers of massive stars can generate very massive stars (VMS), which subsequently evolve to IMBHs \citep[e.g.,][]{PZ2002,PZ2004,PZ2006,Freitag2006,Kremer2020,Gonzalez2021}. 
A VMS with $Z<10^{-3}$ does not undergo pair-instability supernovae and can directly collapse to an IMBH \citep{Spera2017}.

The hierarchical hydrodynamic model of population III (Pop3) star formation from \cite{Sakurai2017} suggested that VMSs can also form in Pop3 star clusters and result in IMBHs.
%No GW signals of IMBH from first run of LIGO \citep{Abbott2017,Abbott2019}.
%In this work, we focus on the IMBH formation and the IMBH-BH GW events from the Pop3 star clusters.
There are a few aspects that suggest Pop3 star clusters might be an environment that contain IMBHs. 
First, the formation times of SMBH seeds and of Pop3 stars overlap. 
Second, the extremely metal-poor gas tends to result in a top-heavy initial mass function \citep[IMF; ][]{Stacy2016,Chon2021,Latif2022}, where a large fraction of Pop3 stars is massive and can either merge to form VMSs or become BHs.
Third, unlike the metal-rich stars, the final mass of the BHs formed from the death of a VMS can be considerably higher without a significant wind mass loss.
Fourth, Pop3 star clusters form inside mini dark matter halos \citep[e.g.,][]{Skinner2020}.
The halos can protect the clusters from tidal disruptions.
Hence, IMBH-BH or BBH mergers can continually occur inside Pop3 star clusters across a wide redshift region.

In previous studies,
\cite{Sakurai2017} investigated the possibility to form VMSs in Pop3 star clusters, but did not follow the long-term evolution of the stars.
\cite{LiuBoyuan2021} studied small Pop3 star clusters, and the dynamical effects are not sufficiently pronounced to form VMSs.
\cite{Reinoso2018} generated a group of low-mass star-cluster models with minimum $Z=0.0001$ to study the collision rate depending on the cluster's initial conditions.
%By performing hydrodynamic simulations, \cite{Skinner2020} investigated the distribution of Pop3 star host halo masses and its relationship to the LW background intensity.
\cite{Tanikawa2022} studied the contribution of BBH mergers from all three populations of stars.

In this work, we investigate the long-term evolution of Pop3 star clusters under mini dark matter halos in detail by performing star-by-star $N$-body simulations with extremely metal-poor stars and IMBH-BH GW mergers.
Section~\ref{sec:method} introduces the numerical method and the initial condition of the $N$-body simulations. 
Section~\ref{sec:result} shows the formation of VMSs and IMBHs, the influence of dark matter halos on the long-term evolution of Pop3 star clusters, and the properties of BBH, IMBH-BH, and IMBH-star mergers.
In Section~\ref{sec:discussion}, we discuss how the uncovered physical mechanisms affect the predictions of GW events. 
In Section~\ref{sec:conclusion}, we draw the conclusion.

\section{Method}
\label{sec:method}

\subsection{PeTar code}
\label{sec:petar} % used for referring to this section from elsewhere
In this work, we used the high-performance $N$-body code \textsc{petar} \citep{Wang2020b} to conduct simulations of Pop3 star clusters.
The code was developed based on the particle-tree and particle-particle algorithm \citep{Oshino2011} under the Framework for Developing Particle Simulator  \citep[\textsc{fdps};][]{Iwasawa2016,Iwasawa2020}.
The slow-down algorithmic regularization method \citep{Wang2020a} was implemented to ensure an accurate and efficient treatment of binary orbital evolution and close encounters.
\textsc{galpy} \citep{Bovy2015} was implemented to address the external potential.

\subsection{bseEmp}
\label{sec:bseemp}

The single and binary stellar evolution codes based on the fast population synthesis method, \textsc{sse} and \textsc{bse}, have been frequently used for $N$-body simulations of star clusters \citep{Hurley2000,Hurley2002,Banerjee2020}.
In this work, we used an extended version, \textsc{bseemp} \citep{Tanikawa2020}, which includes the fitting formula for stars formed in extremely metal-poor environments. 
Thus, we were able to trace the stellar wind mass loss and the BH formation of Pop3 stars with the minimum metallicity, $Z=2\times 10^{-10}$.

The limitations of the fitting formula we adopted are clarified here. 
\cite{Tanikawa2020, Tanikawa2021} constructed the fitting formula, referring to one-dimensional hydrodynamics simulation of $8$ - $1280 M_\odot$ stars by the \textsc{hoshi} code \citep{Takahashi2016, Takahashi2019, Takahashi2018, Yoshida2019}. 
For the high-mass limit, we can correctly follow stellar evolution up to $1280M_\odot$. Moreover, we safely trace the stellar evolution and dynamics of $m>1280M_\odot$ stars, say stars with at least several $10^3 M_\odot$ for the following reason. 
Extrapolating the fitting formula, we find that the maximum radii of stars can be written as $R/10^4 R_\odot=(m/1280M_\odot)^{1.2}$ up to a few $10^4M_\odot$. 
The maximum radii gradually increase with increasing mass. 
Thus, we do not overestimate nor underestimate the merger rate of $m>1280M_\odot$ stars in our simulations. 
At the low-mass limit ($m<8M_\odot$), we approximate Pop3 star evolution as $Z=0.0001$ star evolution as constructed by \cite{Hurley2000}.
As BHs, remnants of $m \gtrsim 20M_\odot$ stars, control the dynamics of the Pop3 star clusters we examined, this approximation has little impact on our results.

\begin{figure}
	% To include a figure from a file named example.*
	% Allowable file formats are eps or ps if compiling using latex
	% or pdf, png, jpg if compiling using pdflatex
	\includegraphics[width=0.9\columnwidth]{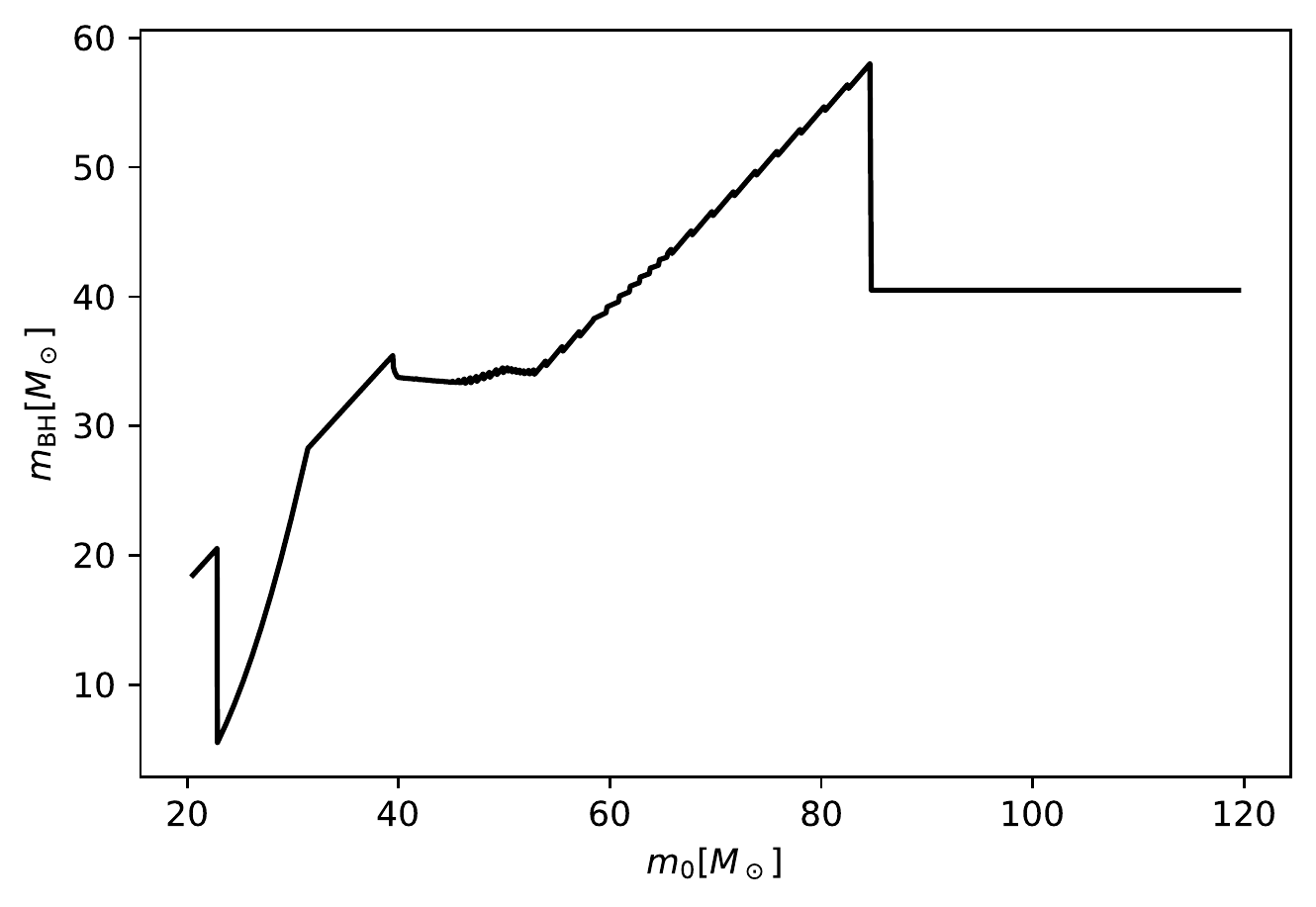}
    \caption{Final BH masses ($\mbh$) vs. zero-age main-sequence masses ($m_0$) for stars with $Z=2\times 10^{-10}$, evolved using \textsc{bseemp}.}
    \label{fig:m0f}
\end{figure}

For supernova model, we adopt the Fryer's rapid model \citep{Fryer2012} with the modification of pair instability supernova \citep{Heger2003} and pulsational pair instability supernova \citep{Woosley2017} modeled by \cite{Belczynski2016} \citep[see also][]{Banerjee2020}. 
Figure \ref{fig:m0f} shows the masses of BHs depending on the zero-age main-sequence masses of Pop3 stars with the upper mass limit of $150~M_\odot$ and $Z=2\times 10^{-10}$ via the pure single stellar evolution. 
We obtained the data by using the standalone code \textsc{bseemp} (single stellar evolution is also included). Stars with $m_0=85$-$121~M_\odot$ (helium core masses of $45$-$65~M_\odot$) cause pulsational pair-instability supernovae, and leave behind approximately $40~M_\odot$ BHs. 
When $m_0>121~M_\odot$ (or helium core masses are $>65~M_\odot$), no BHs form because of pair-instability supernova.

The binary stellar evolution and dynamically driven formation of binaries can produce BHs with masses in the pair-instability mass gap. 
Here, we define the pair-instability mass gap as the mass range from $60$ to $121~M_\odot$, and name BHs within this mass gap as ``pair-instability BHs'' (PIBH). 
First of all, we consider the pair-instability mass gap as the mass range in which no BHs are present if all BH progenitors evolve as single stars. 
The upper bound is the BH mass formed from the minimum naked helium star ($135~M_\odot$) which overcomes a pair-instability supernova explosion and directly collapses to a BH. The lower bound is $\mbh \approx 60 M_\odot$.
This is larger than the generally discussed pair-instability mass gap lower limit ($\sim 40~M_\odot$). Such BHs can be formed from stars with massive hydrogen envelopes and helium cores with mass less than $\sim 45~M_\odot$, because these stars can avoid pulsational pair instability supernovae due to small helium core masses.
%Thus, based on this stellar evolution model of \textsc{bseemp} for Pop3 stars, we expect that BHs with masses between $40-60~M_\odot$ in the pair-instability mass gap can form efficiently.
Thus, based on this stellar evolution model of \textsc{bseemp} for Pop3 stars, we expect that BHs with masses between $40-60~M_\odot$ can form efficiently.

\subsection{Initial conditions}

The hydrodynamic simulations from \cite{Sakurai2017} show that the center of Pop3 star clusters tends to have a high density, which can trigger the hierarchical mergers and VMSs formation subsequently.
For an individual star cluster, the initial density distribution is irregular and is strongly affected by stochastic fluctuations. 
However, the cluster evolves into virial equilibrium state in the crossing time,
\begin{equation}
    \tcr = \sqrt{\frac{\rh^3}{G \Msc}},
\end{equation}
where $\rh$ is the half-mass radius and $\Msc$ is the mass of the star cluster.
After $\tcr$, the morphology of the cluster becomes spherical or elliptical. 
Meanwhile, gas is removed by feedback from massive stars, including radiation, stellar winds and supernovae.
We start our simulation in the virial equilibrium without gas dynamics.
The initial $\Msc$ was fixed to $10^5~M_\odot$, following a similar mass of Model A in \cite{Sakurai2017}.
The initial $\rh = 1$~pc, which is a typical value for observed star clusters.
To cover the different initial central density possibilities, we applied the \cite{michie1963}-\cite{King1966} model where the central concentration can be adjusted by modifying the parameter $W_0$, which indicates the ratio between the core radius ($\rc$) and the tidal radius ($\rt$).
We investigated three values of $W_0$: 6, 9, and 12.
The corresponding $\rc$ values are $0.30$, $0.06$, $0.01$~pc, respectively.
The commonly assumed Plummer model \citep{Plummer1911} for describing the density profile of star clusters is close to the case when $W_0=6$.
When $W_0=12$, the model with a few thousand stars has the highest density concentration. 

The hydrodynamic models of Pop3 star formation suggest a top-heavy IMF \citep{Stacy2016,Chon2021,Latif2022} with the power index being approximately -1.0. 
We assumed a single power-law IMF with the number of stars per mass bin following a linear scale.
\begin{equation}
    \xi (m) \propto m^{-1.0}, \mmin < m < 150 M_\odot. 
\end{equation}
Compared to the canonical IMF with the index of -2.35 \citep[e.g.,][]{Kroupa2001,Chabrier2003}, the top-heavy IMF results in a considerably larger fraction of massive stars.
Thus, we expect the formation of VMSs to be common and a large number of stellar-mass BHs to form.
The cluster eventually evolves into a "dark cluster" \citep{Banerjee2011} where most members are BHs.
Thus, the cluster cannot be detected via electromagnetic observations.
The power index of the IMF has a large uncertainty. 
We simply used -1.0 to represent the effect of the top-heavy IMF.
As the mass range of the IMF is still unclear, we fixed the maximum value to be the classical value of $150~M{_\odot}$.
For $W_0=12$, we investigated three values of $\mmin$: $0.1$, $1$, and $10~M_\odot$.
For other $W_0$, we adopted $\mmin$ to be $1~M_\odot$.

The hydrodynamical models show that massive stars tend to form in the center of dense clouds.
\cite{Plunkett2018} showed that primordial mass-segregation may exist in extremely young embedded clusters.
Thus, we considered the cases with and without primordial mass-segregation for different $W_0$ and $\mmin =1~M_\odot$.
The metallicity of stars was fixed to be $2\times 10^{-10}$.

It is unknown what the real conditions of the mini dark matter halo are where Pop3 star clusters form.
We assumed a \cite{NFW1996} (NFW) dark matter halo with the potential:
\begin{equation}
    \label{eq:nfw}
    \Phi = - \frac{G \Mvir}{r \left [ \log{(1 + \cc}) - \cc/(1+\cc)\right ]} \log{ \left(1+\frac{r}{\rs} \right)},
\end{equation}
where $\Mvir$, $\cc$, and $\rs$ are virial mass, concentration, and scale radius, respectively. 
The virial radius, $\rvir = c\rs$.
To determine the three parameters of the NFW profile, we consider two cases.
In the first case, NFWden, we adopted $\Mvir=4\times 10^7 M_\odot$ and $\rvir = 280$~pc, referring to those in model A from \cite{Sakurai2017}.
We assumed the star clusters formed at redshift, $z=20$. 
The initial concentration is unknown for such redshift.
Based on the recently high-resolution Uchuu cosmological simulation \citep{Ishiyama2021}, for a low-mass dark matter halo ($10^9~M_\odot$), $\cc = 14-15$ at $z=0$, which is similar to that of the present-day Milkyway dark matter halo.
For the Milkyway halo, $\cc = 13.1$ \citep{Gomez2010} or $15.3$ \citep{Bovy2015} depending on models.
With no data of $\cc$ for a lower-mass dark matter halo at $z=20$, we estimate the value as $\cc(z) = \cc/(1+z) \approx 0.728571$, assuming $\cc=15.3$.
In Section~\ref{sec:longterm}, we discuss the influence of $\cc$ on the long-term evolution of Pop3 star clusters.
In this work, we did not include a time-dependent dark matter halo as that of \cite{Sakurai2017}.
We aimed to confirm that the existence of dark matter halos can prevent the disruption of Pop3 star clusters. 
Thus, although the setup of NFWden was an approximation, it was sufficient for the purpose of this work. 

For comparison, we set up another dark matter halo (NFWstd). In this halo, the $\Mvir$ and $\cc$ are the same as those in the NFWden halo, but $\rvir$ is calculated based on the cosmological evolution of dark matter halos described by \cite{Wechsler2002} \cite{Zhao2003}, and \cite{Gomez2010}:
\begin{equation}
    \rvir (z) = \left(  \frac{3 \Mvir(z)}{4\pi \Delta_{\mathrm{vir}}(z) \rhoc (z)} \right)^{1/3},
\end{equation}
with 
\begin{equation}
    \Delta_{\mathrm{vir}}(z) = 18 \pi^2 + 82 \left [  \Omega(z) -1 \right ] - 39 \left [ \Omega(z) -1 \right]^2,
\end{equation}
from \cite{Bryan1998}.
Here the critical density,
\begin{equation}
    \rhoc(z) = \frac{3 H^2(z)}{8 \pi G},
\end{equation}
the normalized mass density of the universe is:
\begin{equation}
    \Omega(z) = \left (\frac{H_0}{H(z)} \right )^2 \Om (1+z)^3,
\end{equation}
and 
\begin{equation}
    H(z) = H_0 \sqrt{\Og (z+1)^4 + \Om (z+1)^3 + \Ol},
\end{equation}
where $H_0$ is the present-day Hubble constant, and $\Og$, $\Om$ and $\Ol$ are the radiation density,  mass density and energy density of the universe at the present day. 
We adopted $H_0 = 70 (\mathrm{km~s}^{-1}) \mathrm{Mpc}^{-1}$, $\Og=0$, $\Om=0.3$, and $\Ol=0.7$.
With the same value of $\Mvir$ as that of NFWden, the estimated $\rvir \approx 522$~pc, approximately twice of that in NFWden.
Thus, the mini dark matter halo of \cite{Sakurai2017} is denser. 

For each combination of $W_0$, mass segregation, and $\mmin$, we performed two models with the NFWstd or the NFWden halos for up to 20 Myr.
As the hierarchical mergers of massive stars occurs in a short time, this is sufficient to investigate the formation of IMBH.
Moreover, we conducted another group of long-term models, where $W_0=6$ and $9$, no mass segregation, and $\mmin=1~M_\odot$ were combined with NFWstd, NFWden, and no dark matter halo (NoDM). 
All models are summarized in Table~\ref{tab:init}. 
For each model, as $N$ was small, we performed 30 simulations with different random seeds for generating the initial masses, positions, and velocities of stars to avoid the stochastic effect.
%\com{AT}{This is just a question. Did you also perform 30 simulations for long-term models?} Yes, for each model we have 30 simulations

% Example table
\begin{table}
	\centering
	\caption{The initial conditions of star clusters. The first column shows the name prefixes of the models, The suffixes '-std', '-den' and '-nodm' refer to the sets of dark matter halos, NFWstd, NFWden and NoDM, respectively. Each model has 30 simulations with different random seeds for generating initial distribution of stars. The second column shows the averaged initial number of stars.}
	\label{tab:init}
	\begin{tabular}{lcccc} % four columns, alignment for each
		\hline
		Name prefix & $\langle N \rangle$ & $W_0$ & Mass segregation & $\mmin$ \\
		\hline
		\multicolumn{5}{c}{Short-term (up to 20 Myr)~~||~  NFWstd (-std) and NFWden (-den)}\\
		\hline
		sw6imf1 & 3369 & 6 & no & 1 \\
		sw9imf1 & 3369 & 9 & no & 1  \\
		sw12imf1 & 3369 & 12 & no & 1  \\
		sw6imf1ms & 3369 & 6 & yes & 1  \\
		sw9imf1ms & 3369 & 9 & yes & 1  \\
		sw12imf1ms & 3369 & 12 & yes & 1 \\
		sw12imf01 & 4882 & 12 & no & 0.1 \\
		sw12imf10 & 1935 & 12 & no & 10  \\
        \hline
        \multicolumn{5}{c}{Long-term (up to 12 Gyr)~~||~ NFWstd, NFWden and NoDM (-nodm)}\\
        \hline
        lw6imf1 & & 6 & no & 1 \\
        lw9imf1 & & 9 & no & 1 \\
        \hline
	\end{tabular}
\end{table}

\section{Results}
\label{sec:result}

\subsection{Formation of VMS and IMBH}
\label{sec:short}

\begin{figure*}
	\includegraphics[width=0.9\textwidth]{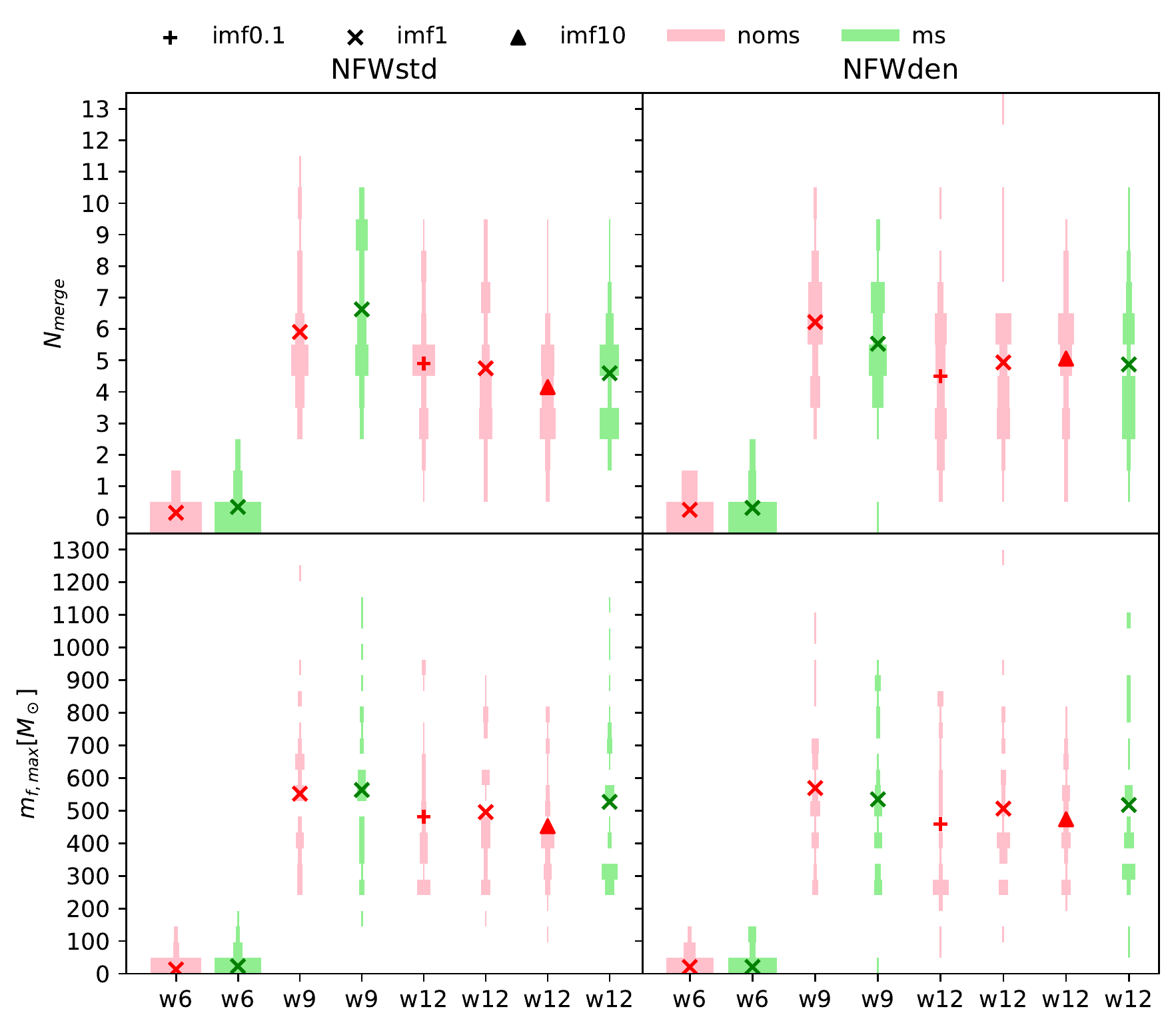}
    \caption{Average numbers of stellar mergers ($\nmerge$) and maximum final masses $\mfmax$ of massive stars for each short-term models. Symbols indicate the average values among the 30 simulations of each model, and the width of boxes represents the number fraction of models within the values of the $y$-axis, similar to a histogram. The bin of the boxes for $\mfmax$ is $50~M_\odot$.
    The colors indicate whether the model has primordial mass segregation; the shapes of symbols represent different $\mmin$.}
    \label{fig:mmerge}
\end{figure*}

We investigated the formation of VMSs via hierarchical mergers.
Figure~\ref{fig:mmerge} shows the number of mergers ($\nmerge$) and the maximum final masses of mergers ($\mfmax$)  (within 20~Myr) for all short-term models.
All mergers with BHs were excluded.
Two distinguishable groups of models can be observed:
the models with $W_0=6$ have almost no mergers and most of their $\mfmax$ are below $50~M_\odot$, whereas other models with $W_0=9$ and 12 have approximately $4-6$ mergers and their $\mfmax$ have a wide distribution from $100$ to $1300~M_\odot$ with an average value approximately $500~M_\odot$.
Thus, the density profile has the largest impact on the merger rate and the formation of VMSs, whereas primordial mass segregation and $\mmin$ have almost no influence.
The distribution of $\mfmax$ for the models where VMSs form displays no clear tendency due to the low number of samples (30 samples per model).

\begin{figure}
	% To include a figure from a file named example.*
	% Allowable file formats are eps or ps if compiling using latex
	% or pdf, png, jpg if compiling using pdflatex
	\includegraphics[width=\columnwidth]{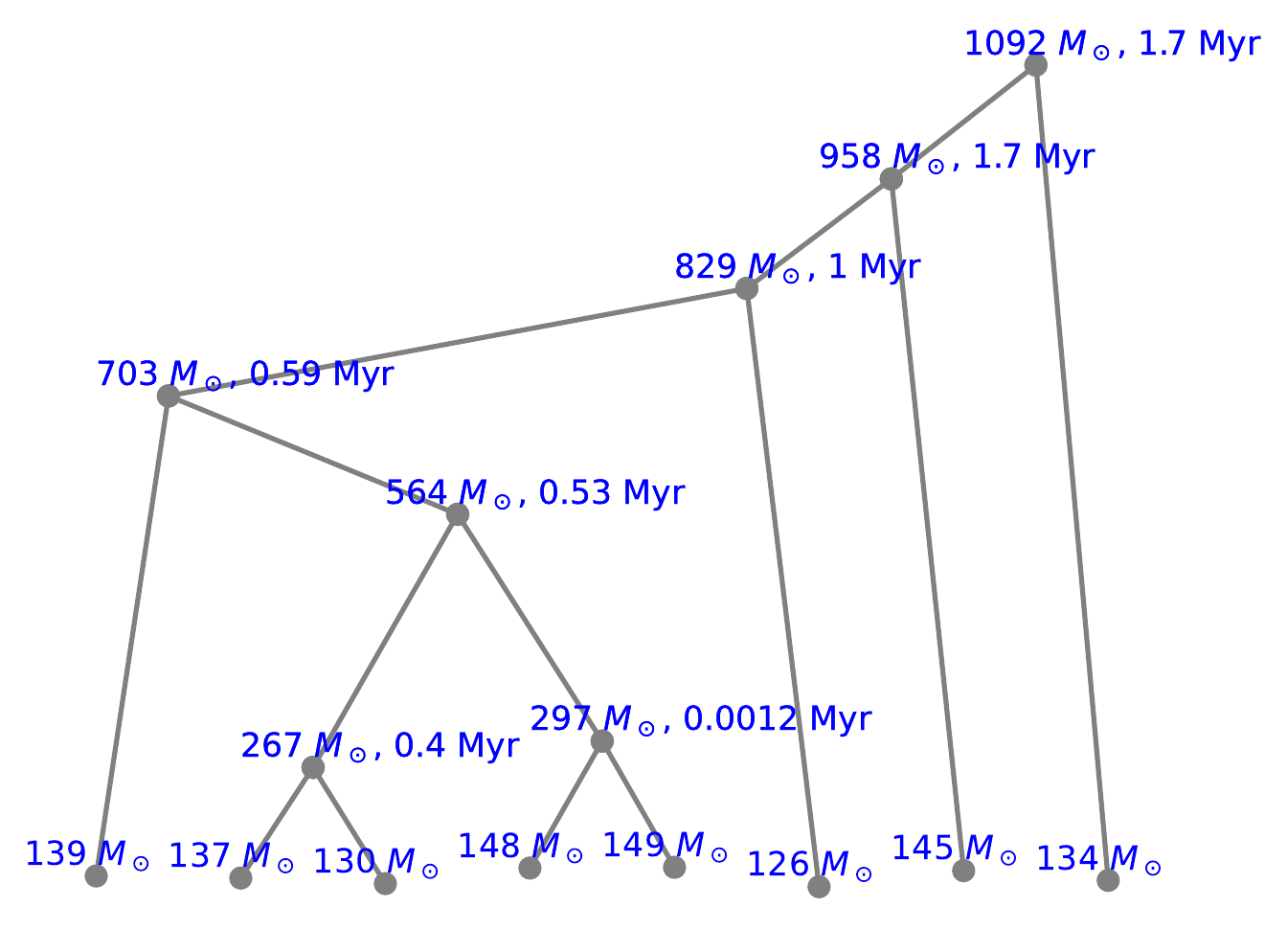}
    \caption{Merger tree for the formation of a VMS in the model msw12m1-den. The masses and merging times of seeds and mergers are  shown. All stars in the merger tree are main-sequence stars.}
    \label{fig:mergetree}
\end{figure}

As a massive star has a short lifetime and strong stellar wind within a few Myr,
the multiple mergers must occur in a short time to successfully produce a VMS.
Figure~\ref{fig:mergetree} shows an example of multiple mergers in one simulation of the model w12imf1ms-den.
Within 2~Myr, a VMS forms after 7 mergers.
The final mass reaches above $1000~M_\odot$.
This is consistent with the prediction of hydrodynamic models from \cite{Sakurai2017}.

To understand how the central density affects the merger rate, we calculated the two-body relaxation time ($\trxl$) within the 10~$\%$ Lagrangian radii, inelastic encounter rate ($\fierl$), and collision rate ($\fcrl$) for different $W_0$ in Table~\ref{tab:cc}.
The sw6imf1, sw9imf1 and sw12imf1 models are used for the calculation. 
All values were evaluated based on the initial condition.

We used the formula for the two-body relaxation time at the half-mass radius in a equal-mass system \citep{Spitzer1987}, but replaced all quantities corresponding to the 10~$\%$ Lagrangian radius ($\rl$):
\begin{equation}
  \trxl = 0.138 \frac{N_{\mathrm{r10}}^{1/2} \rl^{3/2}}{\mrl G^{1/2} \ln \Lambda},
  \label{eq:trh}
\end{equation}
where $\nrl$ and $\mrl$ are the number of stars and the average mass within $\rl$, respectively. 
Here the coefficient 0.138 might not be correct for $\rl$, but we use it as an approximation.

$\fierl$ is calculated by including the gravitational focusing, following Eq.~7.194 of \cite{Binney1987} \citep[see also][]{Hills1976}:
\begin{equation}
     \fierl = 4 \sqrt{\pi} \nrl \sigmarl \left( \rcoll^2 + \frac{G m}{\sigmarl^2} \rcoll \right)
    \label{eq:ie}
\end{equation}
where $\sigmarl$ is the velocity dispersion within $\rl$, and $\rcoll$ is the stellar radius of the star.
Similar to $\trxl$, here we adopted the physical quantities referring to $\rl$ to calculate $\fierl$.

$\fcrl$ represents the collision driven by the dynamically formed binaries via the three-body channel, following the definition in \cite{PZ2002}:
\begin{equation}
    \fcrl = 1\times10^{-3} f_{\mathrm{c}} \frac{\nrl}{\trxl}
    \label{eq:coll}
\end{equation}
where $f_{\mathrm{c}}$ is the effective fraction of dynamically formed binaries that produce a collision. 
We assumed $f_{\mathrm{c}}$ to be 1 for simplicity.
\cite{PZ2002} use the number density and relaxation time at the half-mass radius, but they were the same for different $W_0$. Thus, we used $\rl$ instead to represent the properties close to the cluster center.

Table~\ref{tab:cc} suggests that the inelastic encounter is not the major channel that produces mergers as the rate is extremely low. 
The mergers of dynamically formed binaries well explain our result.
For $W_0=6$, $\fcrl\approx0.48~\mathrm{Myr}^{-1}$, whereas for $W_0=9$ and $12$, $\fcrl\ge1.6~\mathrm{Myr}^{-1}$.
The collision time shown in Figure~\ref{fig:mergetree} confirms this estimation.

We also counted the number of hyperbolic mergers driven by inelastic encounters and dynamically binary mergers for all short-term models.
The results are 115 and 1906, respectively.
Mergers with BHs were excluded.
Thus, the dynamically binary mergers are indeed the major contribution. 

Figure~\ref{fig:mergetree} suggests that the major sources to grow up VMSs are stars with zero-age main-sequence mass above $100~M_\odot$.
Thus, it is reasonable that $\mmin$ has almost no influence on the formation of VMSs.

The mass segregation time is
\begin{equation}
    \tms \approx \frac{\mrl}{m_{\mathrm{ms}}} \trxl,
    \label{eq:tms}
\end{equation}
where $m_{\mathrm{ms}}$ is the mass of the mass-segregated star.
As $\trxl \le 0.22~$Myr for $W_0=6$ and $9$, all stars with mass above $100~M_\odot$ sink into the center shortly after formation.
Thus, the merger rate is also independent of primordial mass segregation.

The dark matter halo has no influence as it has a weak impact on the central potential and affects the collisions within 2~Myr.
We expect that the dark matter halo can trap the escaped stars and can prevent the expansion of the star cluster due to long-term dynamical evolution, driven by the central binary heating. 
We discuss this in the next section.

\begin{table*}
    \centering
    \caption{Number of stars ($\nrl$), average density ($\rhorl)$,  relaxation time ($\trxl$),  inelastic encounter rate ($\fierl$) and  collision rate ($\fcrl$) calculated within the $10~\%$ Lagrangian radii for the initial conditions of sw6imf1, sw9imf1, and sw12imf1 models. The errors are the standard derivation evaluated from the 30 simulations of each model.}
    \begin{tabular}{ccccccc}
        \hline
       Model &   $W_0$ & $\nrl[\mathrm{pc}^{-3}]$ & $\rhorl [M_\odot \mathrm{pc}^{-3}]$ & $\trxl [\mathrm{Myr}^{-1}]$ & $\fierl[\mathrm{Myr}^{-1}]$ & $\fcrl[\mathrm{Myr}^{-1}]$ \\
        \hline
        sw6imf1 & 6 & 343 $\pm$ 24 & 6.3e+04 $\pm$ 9.4e+03 & 0.72 $\pm$ 0.072 & 3.01e-05 $\pm$ 3.91e-06 & 0.48 $\pm$ 0.047 \\
        sw9imf1 & 9 & 331 $\pm$ 15 & 1e+06 $\pm$ 1.9e+05 & 0.18 $\pm$ 0.018 & 3.92e-04 $\pm$ 7.18e-05 & 1.8 $\pm$ 0.2 \\
        sw12imf1 & 12 & 322 $\pm$ 22 & 6.9e+05 $\pm$ 2.5e+05 & 0.22 $\pm$ 0.037 & 2.62e-04 $\pm$ 8.80e-05 & 1.6 $\pm$ 0.31 \\
        \hline
    \end{tabular}
    \label{tab:cc}
\end{table*}

\begin{figure}
	\includegraphics[width=\columnwidth]{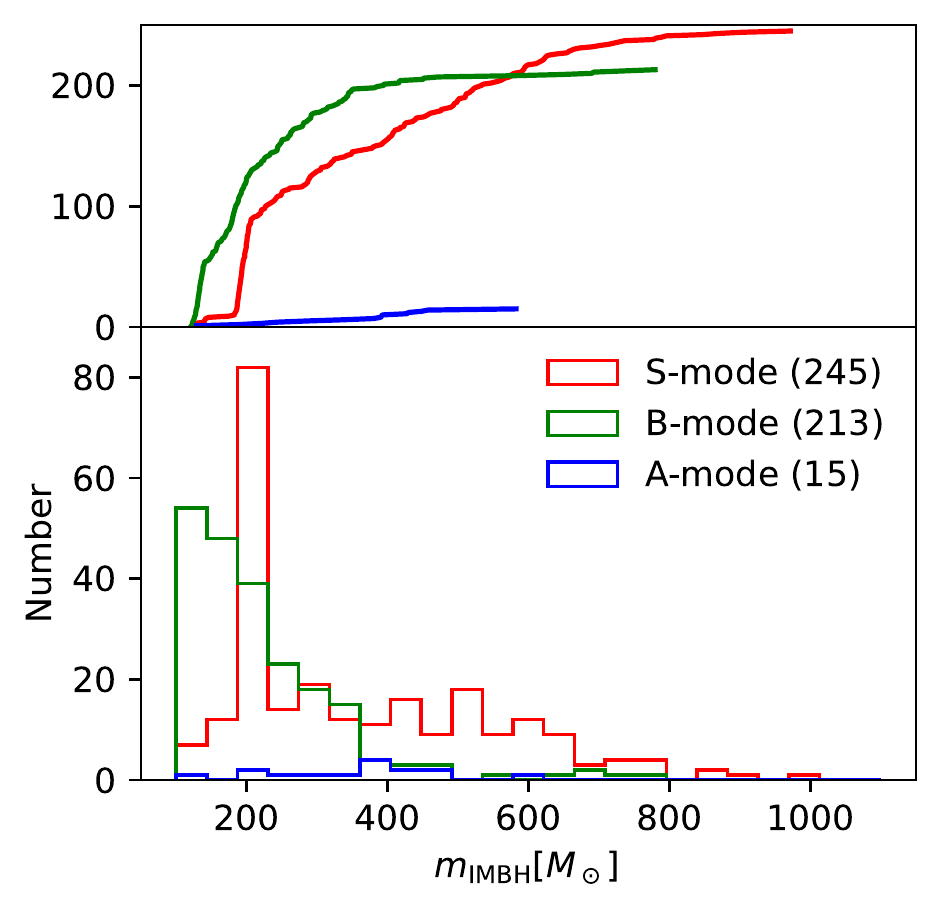}
    \caption{Mass spectrum of IMBHs from three formation modes. The upper panel shows the cumulative distribution, and the bottom panel shows the histogram. All short-term models with $W_0=9$ and 12 are included. The minimum mass limit of IMBH is 121~$M_\odot$, above the pair-instability supernovae mass gap. The number of formed IMBHs for each mode is shown in the legend.}
    \label{fig:mfshort}
\end{figure}

Once a VMS forms, an IMBH can form via three modes:
\begin{itemize}
    \item S-mode: a VMS directly evolves to an IMBH as a result of stellar evolution. For example, the VMS shown in Figure~\ref{fig:mergetree} evolves to a first asymptotic giant branch star (FAGB) with $877~M_\odot$. Then, it becomes an IMBH with 789~$M_\odot$ at $3.17$~Myr. The wind mass loss reduces its mass during stellar evolution. 
    \item B-mode: a VMS evolves to an IMBH in a binary system. Mass transfer may occur between the two binary members. The VMS evolves to a main-sequence naked helium star (HeMS), and then becomes an IMBH.
    \item A-mode:  a VMS can be accreted by a low-mass BH in a VMS-BH binary system and eventually form an low-mass IMBH. During this process, a large fraction (roughly half) of the VMS mass is lost. 
\end{itemize}
$\mfmax$ shows no intrinsic difference among models with $W_0=9$ and 12.
Thus, we collected all data of formed IMBHs in all these models for better statistical analysis.
Figure~\ref{fig:mfshort} shows the mass spectrum of formed IMBHs from the three modes.
We set the minimum mass limit for IMBHs to be 121~$M_\odot$, which is the upper limit of the pair-instability supernovae mass gap.
The S-mode and the B-mode contribute equally to the number of IMBHs, while a few IMBHs form from the A-mode. 
The S-mode produces more massive IMBHs than the B-mode.
There is a peak around $200~M_\odot$ in the S-mode, mostly from a massive star undergoing only one merger, which more commonly occurs compared to hierarchical mergers.
The A-mode only contributes to a few IMBHs with masses below $500~M_\odot$, except one case of $582~M_\odot$. 

\begin{table}
    \centering
    \caption{IMBH-VMS mergers in short-term models. Columns show merge times ($\tf$), semi-major axes ($a$), eccentricities ($e$), masses of primary members (IMBHs; $\mimbh$), masses of secondary members (VMSs; $\mvms$), and the final masses of IMBHs ($\mf$), respectively.}
    \begin{tabular}{cccccc}
       \hline
        $\tf$ [Myr] & $a [R_\odot]$ & e & $\mimbh [M_\odot]$ & $\mvms [M_\odot]$ & $\mf [M_\odot]$\\
       \hline
       2.25 &      2747.1 &    0.99546 &        250 &        413 &             457\\
       2.74 &      1906.5 &  3.389e-07 &        166 &        833 &             582\\
       2.54 &     10980.9 &    0.99756 &        131 &        415 &             338\\
       \hline
    \end{tabular}
    \label{tab:imbh-vms}
\end{table}

Three events in the A-mode are IMBH-VMS mergers, where the masses of the progenitor BHs are already above $121~M_\odot$. 
Table~\ref{tab:imbh-vms} shows the data for the three IMBH-VMS mergers.
The formed IMBH with $582~M_\odot$ had the progenitor VMS of $833~M_\odot$.
The binary is circularized before merging. 
Other two mergers have high eccentric orbits.

\subsection{Long-term evolution}
\label{sec:longterm}

\begin{figure*}
	% To include a figure from a file named example.*
	% Allowable file formats are eps or ps if compiling using latex
	% or pdf, png, jpg if compiling using pdflatex
	\includegraphics[width=0.9\textwidth]{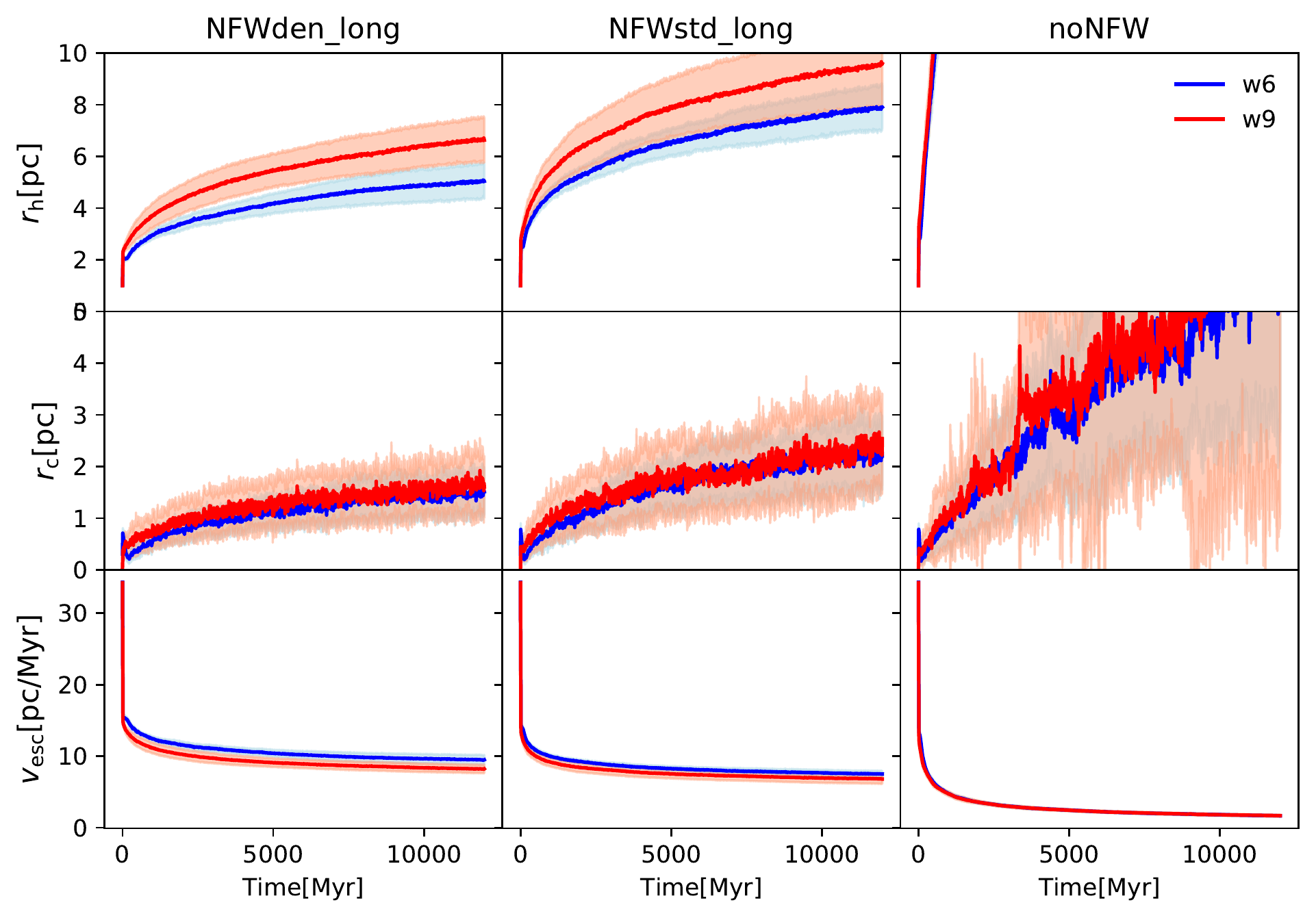}
    \caption{Evolution of the half-mass radii ($\rh$), core radius ($\rc$), and escape velocity only from stars ($\vescsc$) for the long-term models. The solid curves are the average of 30 simulations for each model. The area of light colors indicates the standard derivative.}
    \label{fig:rhrc}
\end{figure*}

The short-term models suggest that IMBHs form at early times due to the formation of VMSs.
%We focus on the formation of IMBHs for the short-term models.
Hereafter we analyze the long-term evolution of Pop3 star clusters and the GW mergers under the influence of mini dark matter halos.
Figure~\ref{fig:rhrc} compares the evolution of $\rh$, $\rc$, and escape velocity ($\vescsc$) for the long-term models. 
Here $\vescsc$ is estimated by only including the gravitational potential from stars.
Without a dark matter halo, the NoDM clusters expand fast for the first 1000~Myr. 
$\rc$ reaches approximately 5~pc at 12~Gyr.
In contrast, models with dark matter halos can keep $\rh$ within 10~pc during the long-term evolution.
The strong stellar winds from massive stars cause an fast expansion for the first 100~Myr, which also reduces $\vescsc$ significantly. 

Stars escape from clusters via two mechanisms: tidal evaporation and ejection by binaries in the cluster centre. 
When the clusters evolve in the galactic potential (which is not included in our models), stars in the halo of clusters feel strong tidal forces and continue escaping. 
Thus, the noDM models are difficult to survive due to tidal evaporation.
However, when a dark matter halo exists, its strong gravitational potential significantly increases the escape velocity of stars; thus, the tidal evaporation process is suppressed. 
Models with dark matter halos in our simulations can survive till today due to the "protection" of dark matter halo.
Thus, we expect that the mergers between stellar mass BHs and IMBHs can continue until today in the long-existed Pop3 star clusters.

Although we only included two models of dark matter halos in our simulations, we can estimate the effect of NFW dark matter halos in general by evaluating the additional central escape velocity ($\vescnfw$) provided by the halo.
Based on Equation~\ref{eq:nfw}, 
\begin{equation}
    \begin{aligned}
    \vescnfw = & \sqrt{\left[\Phi\left(\infty \right) - \Phi \left( 0 \right) \right] } \\
        = & \sqrt{\frac{G \Mvir}{\rs \left[ \log{(1+\cc}) - \cc/(1+\cc) \right ]}}.
    \end{aligned}
    \label{eq:vescnfw}
\end{equation}
For the NFWden and NFWstd halos, the corresponding values are approximately 61 and 45 pc~Myr$^{-1}$, respectively.
$\vescsc$ from star clusters are below 15~pc~Myr$^{-1}$ as shown in Figure~\ref{fig:rhrc}.
As $\vescnfw$ is more than twice $\vescsc$, stars are difficult to escape under the dark matter halos.

\begin{figure}
	% To include a figure from a file named example.*
	% Allowable file formats are eps or ps if compiling using latex
	% or pdf, png, jpg if compiling using pdflatex
	\includegraphics[width=0.9\columnwidth]{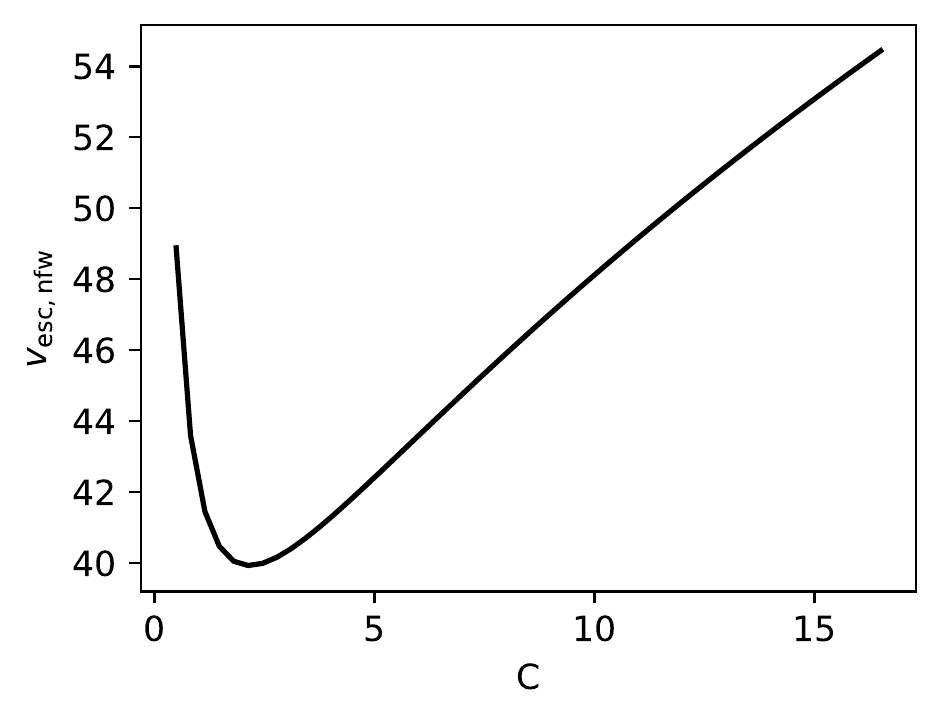}
    \caption{Central escape velocity depending on the concentration of NFW profile. $\Mvir$ and $\rvir$ are chosen from the NFWstd halo.}
    \label{fig:vescnfw}
\end{figure}

$\vescnfw$ depends on $\Mvir$ and $\rs$ in a simple manner.
The dependence on $\cc$ is not intuitive.
Figure~\ref{fig:vescnfw} shows how $\vescnfw$ depends on $\cc$ when $\Mvir$ and $\rvir$ are fixed ($4\times 10^7 M_\odot$ and $522$~pc referring to values of the NFWstd halo, respectively).
When $\cc$ increases from 0.5 to 16.5, $\vescnfw$ decreases and then increases.
The minimum value ($40$ pc~Myr$^{-1}$) appears at $\cc \approx 2.13$.
This value is slightly below that of the NFWstd halo in our simulations and is considerably larger than $\vescsc$ of star clusters. 
Thus, our result is not strongly affected by the choice of $\cc$.

The total escape velocity ($\vesc$) affects the GW merger rate.
The orbit of a tight binary in the center of a star cluster shrinks via three-body interactions.
However, when $a$ becomes smaller, the center-of-the-mass of the binary gain a larger kinetic energy.
Finally the binary is ejected if not merged.
The minimum $a$ that triggers the ejection depends on $\vesc$ as \citep{Miller2002,Mapelli2021}:
\begin{equation}
    a_{\mathrm{ej}} \propto \frac{\langle m \rangle^2 m_1 m_2}{\left(m_1+m_2\right)^3 \vesc^2}
\end{equation}
where $m_1$, $m_2$, and $\langle m \rangle$ are the masses of two components of the binary and the average stellar mass of the cluster, respectively.
As the GW merger timescale strongly depends on $a$, a larger $\vesc$ results in a higher possibility of GW mergers inside clusters.
Therefore, dark matter halos help to prevent the disruption of Pop3 star clusters and also increases the merger rate of GW.

\subsection{IMBH-BH mergers}

\begin{figure}
	% To include a figure from a file named example.*
	% Allowable file formats are eps or ps if compiling using latex
	% or pdf, png, jpg if compiling using pdflatex
	\includegraphics[width=\columnwidth]{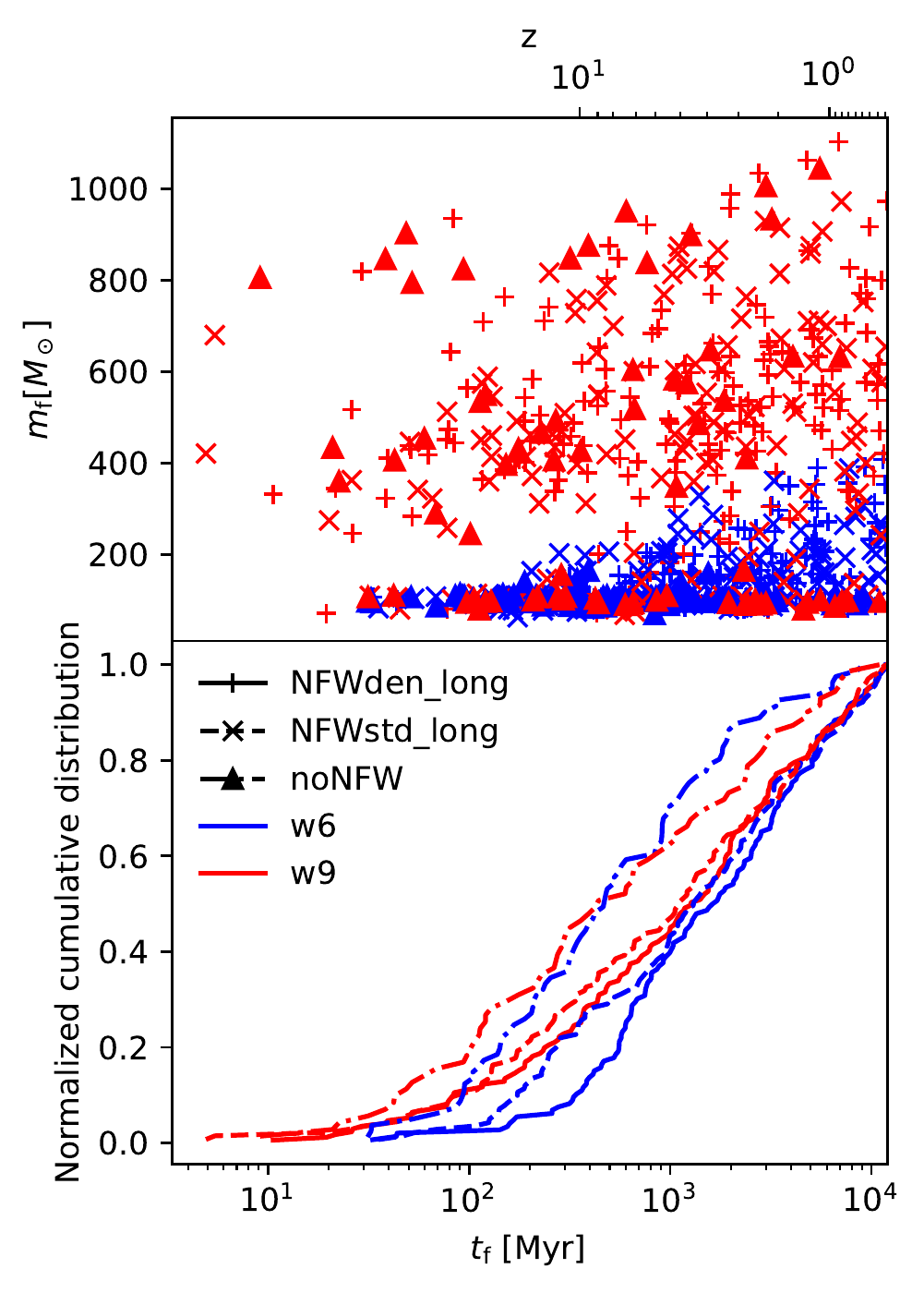}
    \caption{Upper panel: $\mf$ of each merger of two BHs or IMBHs for all long-term models. Lower panel: the normalized cumulative distribution of the merger time ($\tf$). The corresponding redshift ($z$) is shown in the upper $x$-axis.}
    \label{fig:mflong}
\end{figure}

Due to the top-heavy IMF, most members in the survival star clusters under the dark matter halo are BHs. 
Thus, even the clusters may still exist today in the Galaxy, they cannot  be detected by the electromagnetic observations. 
This is similar to the "dark clusters" discussed in \cite{Banerjee2011}.
If the clusters have the high central density like that from \cite{Sakurai2017} and the models with $W_0=9$ and 12, we expect that the existing IMBH may interact with stellar-mass BH and generate GW signals.
The upper panel of Figure~\ref{fig:mflong} shows $\mf$ of all mergers of BHs and IMBHs for the long-term models.
The lower panel shows the normalized cumulative distribution of merger time.
A large fraction of mergers ($0.4-0.8$) occurs at the first 1~Gyr evolution for all models,  because of a high central density. 
As $\rc$ increases (see Figure~\ref{fig:rhrc}), the merger rate decreases, but mergers continue until the end of simulations (12~Gyr).
This suggests that IMBH-BH and BBH mergers can be detected at all redshift regions.
The merger time distribution follows the standard distribution in which $\propto t^{-1}$ \citep[e.g.,][]{Totani2008}.
% by the space borne GW detectors like LISA, TianQin, Taiji, and DECIGO.
%The signals, especially the BBH mergers, may also be captured by the ground-based detectors \citep[e.g.,][]{Jani2020}. 
%\com{AT}{Are the merger time distribution $\propto t^{-1}$? It is informative to compare your distribution with the standard distribution $\propto t^{-1}$.}

\begin{figure}
	% To include a figure from a file named example.*
	% Allowable file formats are eps or ps if compiling using latex
	% or pdf, png, jpg if compiling using pdflatex
	\includegraphics[width=\columnwidth]{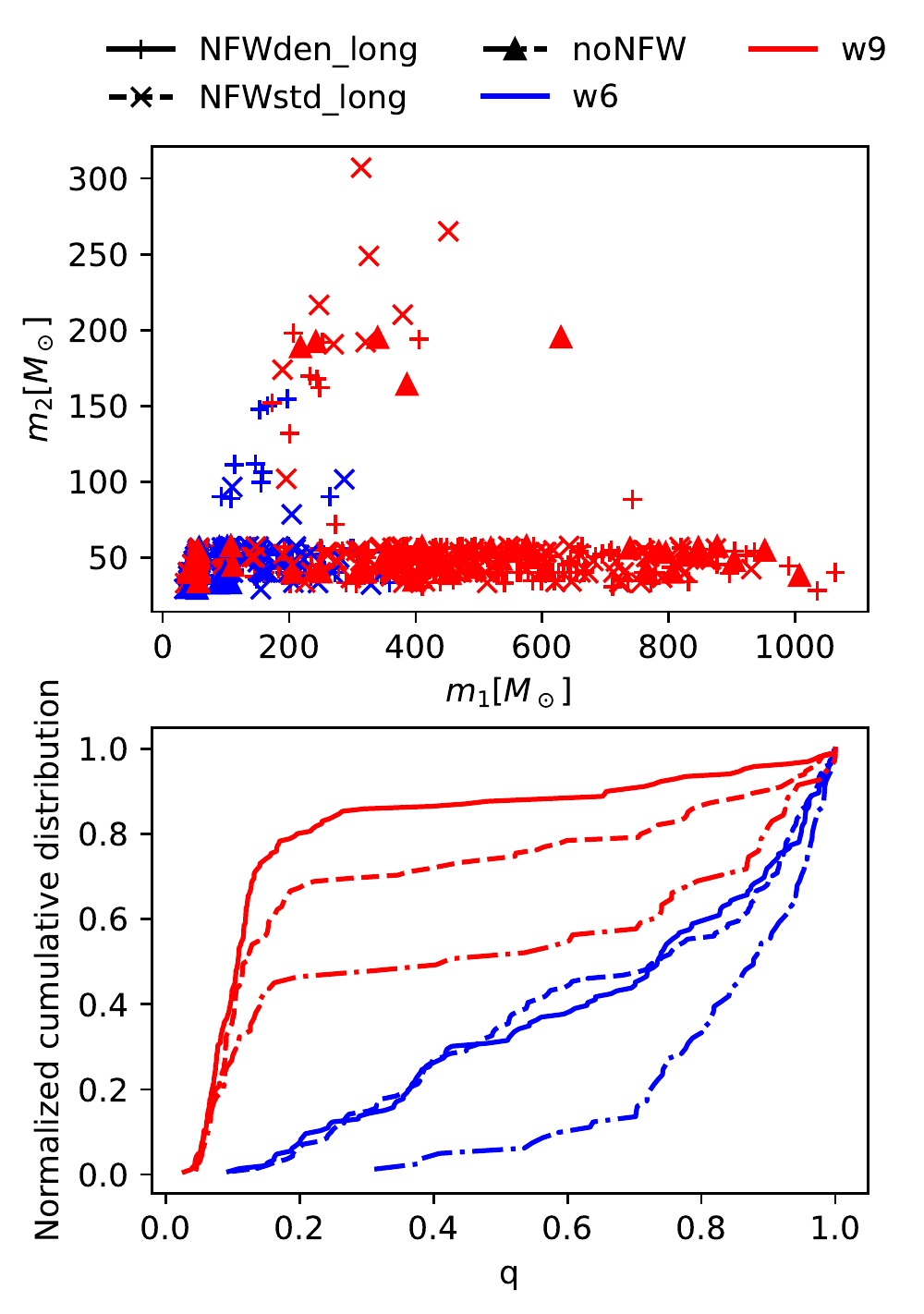}
    \caption{Upper panel: $m_1$ v.s. $m_2$ of each merger of two BHs or IMBHs for all long-term models. Lower panel: normalized cumulative distribution of the mass ratio ($q$).}
    \label{fig:qlong}
\end{figure}

Figure~\ref{fig:qlong} shows the $m_1$ and $m_2$ of each merger and the normalized cumulative distribution of the mass ratio $q$.
With IMBHs ($W_0=9$), a large fraction of mergers ($0.4-0.8$) has $q<0.2$, which is larger than the definition of intermediate mass-ratio inspirals (IMRI) discussed in general ($10^{-5} - 10^{-2}$).
A denser dark matter halo results in more low-$q$ mergers.
For models without IMBHs, the dark matter halo also helps to increase the fraction of low-$q$ mergers.
Few massive BBH mergers have binary members with a mass above $200~M_\odot$ each.
The final masses can be above 100~$M_\odot$, larger than most detected GW events from LVK \citep{LVK2021}.

We did not find IMBH formation for $W_0=6$ even during the long-term evolution. 
An IMBH may form via hierarchical mergers of BHs, like the extremely dense GC \citep{Giersz2015} and galactic center \citep{Antonini2019,Kroupa2020}.
However, the density in the Pop3 model is not sufficient to trigger on this channel. 
%This is because many stellar-mass BHs (so-called BH subsystem) prevent a single IMBH formation. 
%This is consistent with the argument by \cite{Giersz2015} in which a GC can bring up an IMBH a few Gyr after the GC formation only when its BH subsystem disappears due to dynamical ejection.

%\begin{figure}
%	% To include a figure from a file named example.*
%	% Allowable file formats are eps or ps if compiling using latex
%	% or pdf, png, jpg if compiling using pdflatex
%	\includegraphics[width=\columnwidth]{mergebhlong.pdf}
%    \caption{The upper panel: $\mf$ and $m_1$ v.s. $m_2$ of each merger for all long-term %models. The lower panel: the normalized cumulative distribution of the merger time %(left) and of the mass ratio ($q$; right).}
%    \label{fig:mergebhlong}
%\end{figure}

\begin{figure}
	% To include a figure from a file named example.*
	% Allowable file formats are eps or ps if compiling using latex
	% or pdf, png, jpg if compiling using pdflatex
	\includegraphics[width=0.9\columnwidth]{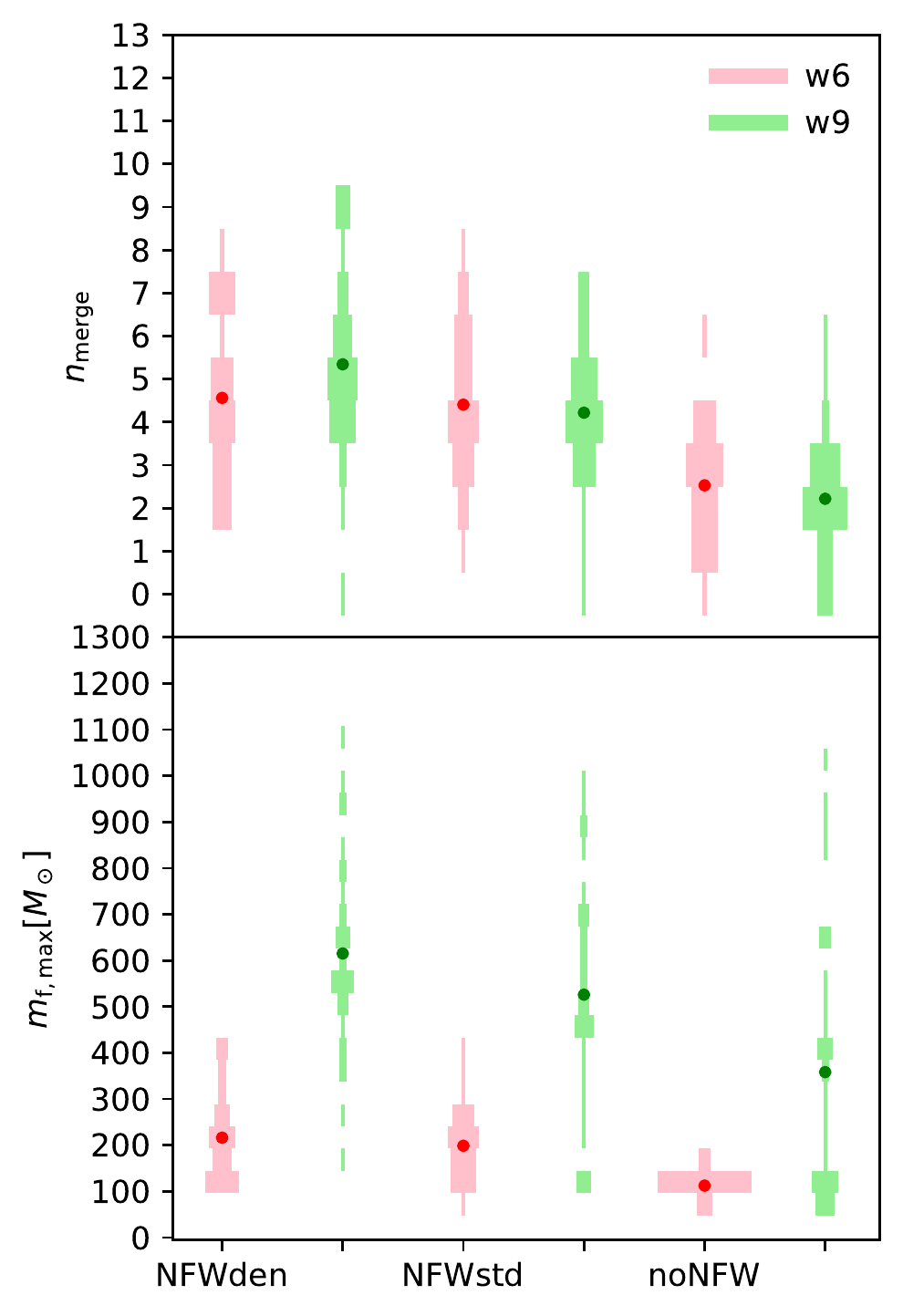}
    \caption{Average number of mergers and the final maximum masses of BHs for the long-term models. The plotting style is similar to that of Figure~\ref{fig:mmerge}.}
    \label{fig:mergebhavelong}
\end{figure}

Figure~\ref{fig:mergebhavelong} shows the average values and the standard derivations of $\nmerge$, $\mfmax$, and the mass ratios ($q$) of BBH mergers (including IMBHs) for the long-term models.
As observed, a higher density of dark matter halo results in a larger average $\nmerge$ and a larger average $\mfmax$. 
This is expected as the collision rate is higher for denser clusters under a denser dark matter halo.
Our models without dark matter halo can also generate IMBH-BH mergers.
But this is unlikely to occur in the real galactic environment, as these star clusters  easily suffer tidal disruption.

\subsection{Mergers in pair-instability mass gap}

%We also count the mergers with BHs in the mass gap ranging from $60$ to $121~M_\odot$ (named as PIBH), which should originate from binary stellar evolution or BBH mergers.
We also counted the mergers with PIBHs, which should originate from binary stellar evolution or BBH mergers.
Table~\ref{tab:pibh} shows the average number of mergers including PIBHs.
The models with $W_0=6$ have a systematically higher number of mergers with PIBHs compared to that of the models with $W_0=9$.
Thus, IMBHs can suppress the formation of PIBHs.
There are two possible reasons:
the formation of VMSs can reduce the number of massive-star binaries that can form PIBHs; and the existence of IMBHs suppresses the formation of low-mass binaries with PIBHs.
Among all merger types, PIBH-LBH contributes the most.

\begin{table}
    \centering
    \caption{Average number of mergers including pair instability mass gap BH (PIBH)  for the long-term models. BPIBH indicates that both binary members are PIBHs; LBH is low-mass BH ($<60~M_\odot$).}
    \begin{tabular}{ccccc}
        \hline
        Halo & King & IMBH-PIBH & BPIBH & PIBH-LBH \\
        \hline
  %      NFWden & w6 & 0.75 & 2.16 & 1.25 \\
  %      NFWden & w9 & 2.75 & 0.22 & 0.12 \\
  %      NFWstd & w6 & 0.66 & 2.41 & 0.72 \\
  %      NFWstd & w9 & 2.12 & 0.59 & 0.22 \\
  %      noNFW & w6 & 0.00 & 1.81 & 0.66 \\
  %      noNFW & w9 & 0.75 & 0.69 & 0.28 \\
        NFWden & w6 & 0.12 & 0.094 & 0.88 \\
        NFWden & w9 & 0.062 & 0 & 0.094 \\
        NFWstd & w6 & 0.062 & 0.031 & 0.88 \\
        NFWstd & w9 & 0.031 & 0 & 0.16 \\
        noNFW & w6 & 0 & 0 & 0.22 \\
        noNFW & w9 & 0 & 0 & 0.062 \\
        \hline
    \end{tabular}
    \label{tab:pibh}
\end{table}

The presence of mergers with PIBHs is a striking feature of BH mergers in Pop3 star clusters. Isolated Pop3 binary stars cannot form such BH mergers \citep{Hijikawa2021}, unless we assume inefficient convective overshoot in Pop3 single star evolution \citep{Tanikawa2021MNRAS}, or detail modeling of mass transfer in Pop3 binary stars \citep{Kinugawa2021}. \cite{LiuBoyuan2020ApJL} shows that Pop3 star clusters can form PIBH-LBH like GW190521 \citep{Abbott2020}. 
In contrast, Table \ref{tab:pibh} clearly shows that Pop3 star clusters can be formation sites of a wide variety of mergers with PIBHs including IMBH-PIBHs and BPIBHs, independently of Pop3 single and binary star evolution.

GW190521 is considered to contain at least one PIBH \citep{Abbott2020}, although it can be an IMBH-LBH merger \citep{Fishbach2020, Nitz2021}. 
From Figure~\ref{fig:qlong}, various mass combinations of BH mergers with the total mass of approximately $150~M_\odot$ can be observed. If we regard PIBH-LBH mergers as GW190521-like mergers, the PIBH-LBH merger rate is $0.01 - 0.15~\mathrm{yr}^{-1}~\mathrm{Gpc}^{-3}$, which is comparable to the inferred merger rate of GW190521-like mergers \citep{Abbott2020ApJL}. Thus, Pop3 star clusters could reproduce GW190521. Here, we have to remark that GW190521 can be formed not only in Pop3 star clusters, but also in isolated binary stars \citep{Belczynski2020, Kinugawa2021, Tanikawa2021MNRAS}, young star clusters \citep{DiCarlo2020, Kremer2020, ArcaSedda2021, Gonzalez2021, Rizzuto2021}, globular clusters \citep{Rodriguez2019}, and galactic centers \citep{Tagawa2021, Fragione2022arXiv}.

%\com{AT}{Are there pair instability mass gap events ($40-130M_\odot$ BH mergers)? I expect that the mass gap should be buried because of several mechanisms: no (or weak) stellar wind, BH-BH mergers, and PMS-MS mergers. I'd like to know which mechanism is dominant. I think it is also useful to compare your results with Hijikawa et al. (2021) in which there is no mass gap event.}

\subsection{Peak frequency and characteristic strain for GW detection}

\begin{figure}
	% To include a figure from a file named example.*
	% Allowable file formats are eps or ps if compiling using latex
	% or pdf, png, jpg if compiling using pdflatex
	\includegraphics[width=\columnwidth]{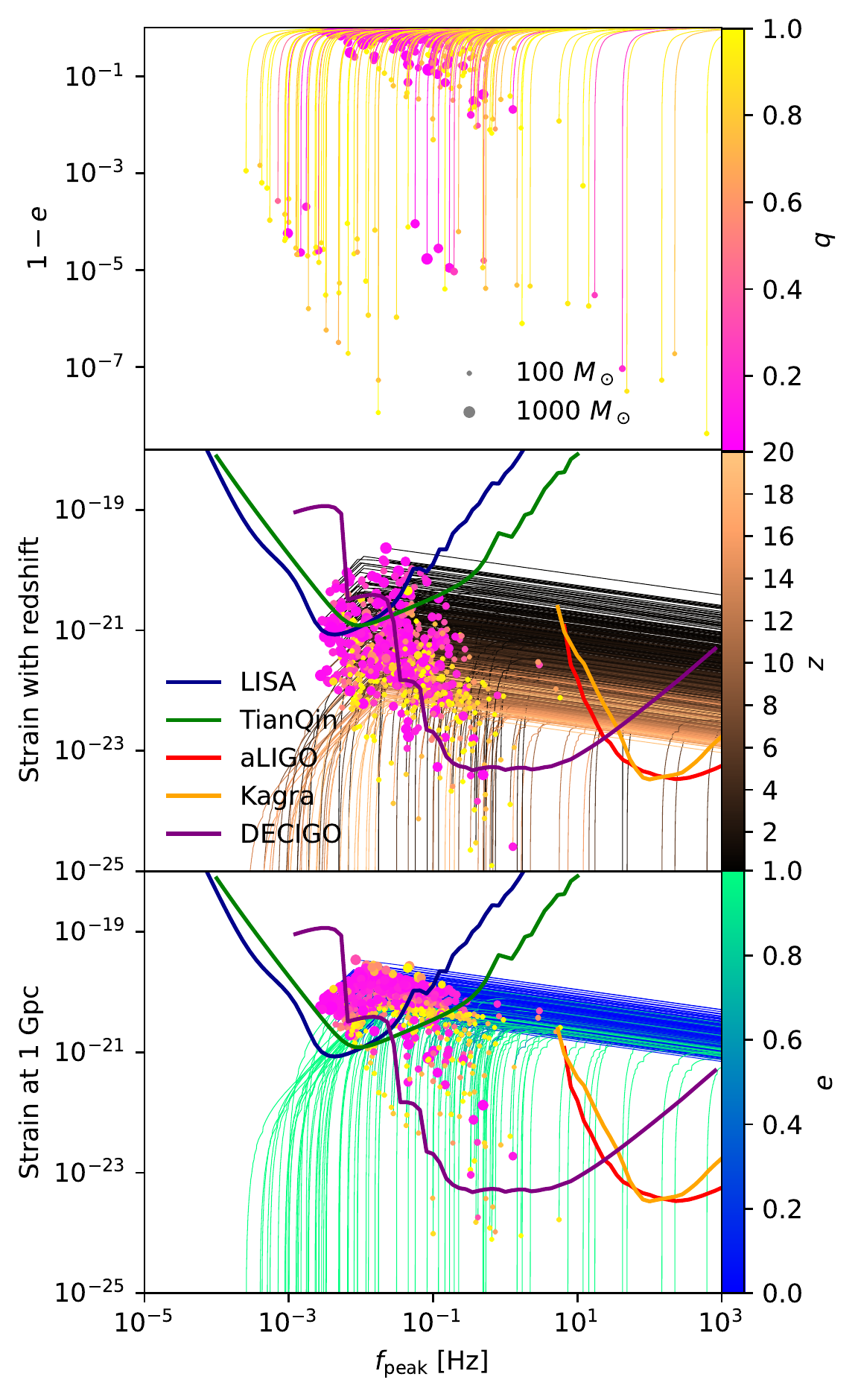}
    \caption{Eccentricity, peak frequency of GW signals ($\fpeak$) and, characteristic strain of all IMBH-BH and BBH mergers in the long-term models. Symbols show the initial condition of each binary and curves show their evolution due to the GW radiation. Sizes of symbols indicate the total masses of binaries; colors indicate the mass ratios ($q$) for the upper panel, redshift ($z$) for the central panel, and $e$ (curve colors) for the lower panel. $z$ with corresponding Hubble distance is used to evaluate strain for the central panel, whereas a fixed distance of 1~Gpc with $z \approx 0.233$ is assumed for the strain in the lower panel. The sensitivity curves of 5 GW detectors are shown.}
    \label{fig:obs}
\end{figure}

The IMBH-BH and BBH mergers in the Pop3 star clusters can be detected by the future space-borne GW detectors, like LISA, TianQin, and Taiji.
We calculated the peak frequency ($\fpeak$) of GW detection and characteristic strain of our mergers.
We adopted the estimation of $\fpeak$ from \cite{Hamers2021}, which is an improved version for low eccentric binaries based on \cite{Wen2003}.
The detail formula is:
\begin{equation}
\begin{aligned}
    \fpeak =  & \frac{\sqrt{G \left( m_1+m_2 \right)}}{\pi} \\
    & \frac{1 - 1.01678e + 5.57372e^2 - 4.9271e^3 + 1.68506e^4}{ \left[ a \left ( 1-e^2 \right ) \right ]^{1.5}}.
\end{aligned}
\end{equation}

The upper panel of Figure~\ref{fig:obs} shows $1-e$ v.s. $\fpeak$. 
The evolution curves are integrated using $\adot$ and $\edot$ based on the secular change estimation from \cite{Peters1964}:
\begin{equation}
    \begin{aligned}
    \adot = & - \frac{\beta F(e)}{a^3}, \\
    \edot = & - \frac{19}{12}  \frac{\beta e}{a^4 \left ( 1 - e^2 \right)^{5/2}} \left ( 1 + \frac{121}{304} e^2 \right), \\
    \end{aligned}
\end{equation}
where the factor,
\begin{equation}
    \begin{aligned}
    F(e) = & \sum_{n=1}^{\infty} g(n,e) \\ 
         = & \frac{1}{\left ( 1 - e^2 \right)^{7/2}} \left ( 1 + \frac{73}{24} e^2 + \frac{37}{96} e^4 \right), \\
    \beta = & \frac{64}{5} \frac{G^3 m_1 m_2 \left ( m_1 + m_2 \right )}{\vc^5}.
    \end{aligned}
\end{equation}
Here, $g(n,e)$ is Equation 20 from \cite{Peters1963}, and $\vc$ is the speed of light.
The integration stops when $e$ approaches zero.

We calculated the characteristic strain ($\hcn$) corresponding to $\fpeak$ by using the formula from \cite{Kremer2019}:
\begin{equation}
    \hcn^2 = \frac{2}{3 \pi^{4/3}} \frac{G^{5/3}}{\vc^3} \frac{M_{\mathrm{c},z}^{5/3}}{D^2} \frac{1}{\fnz^{1/3} \left(1+z\right)^2} \frac{2}{n}^{2/3} \frac{g(n,e)}{F(e)},
\end{equation}
where $M_{\mathrm{c},z}$ is the observed chirp mass with the redshift $z$, and $\fnz$ is the observed frequency of the $n$th harmonic.
$n$ can be calculated by $\fpeak/\forb$, where $\forb$ is the rest-frame frequency of a binary. 
With the limited observation time, $T_{\mathrm{obs}}$, $\hcn$ is multiplied by the square root of $\mathrm{min}[1, \dot{f_{n}} (T_{\mathrm{obs}}/f_{n})]$ where $f_{n}$ is the rest-frame frequency with $n$th harmonic ($f_{n} = \fpeak$ in our calculation).
We adopted that $T_{\mathrm{obs}}= 4~$years.

We found a group of extremely eccentric IMBH-BH and BBH mergers with $1-e<10^{-5}$.
Some even have $1-e<10^{-7}$.
Their $\fpeak$ spreads in a large region.

The central panel of Figure~\ref{fig:obs} shows the characteristic strain of each merger vs. $\fpeak$ considering $z$ and the corresponding Hubble distances. 
Their $z$ are estimated by $\tf$ as shown in Figure~\ref{fig:mflong}.
The hubble distances are calculated by $c z / H_0$.
We also plot the space-borne GW detectors, LISA, TianQin and DECIGO, sensitivity curves \citep{Robson2019,TQ2019,Kawamura2011}, and the ground-base detectors, advanced LIGO (aLIGO) and Kagra \citep{Miller2015,Michimura2020}.
The Taiji mission has a similar planned arm length to that of LISA; thus, we expect its sensitivity curve to be close to that of LISA.
For LISA, TianQin, and Taiji, a part of low-$z$ mergers can be detected.
Most mergers can be covered by DECIGO, aLIGO, and Kagra. 
In simulations, the initial $a$ and $e$ of mergers are not recorded at the times when the binaries form, but at the times slightly earlier than $\tf$ (depending on the integration time step sizes).
Thus, the starting points of some mergers are already inside the detectable region.

The lower panel of Figure~\ref{fig:obs} shows the strain of mergers assuming all events occur at 1~Gpc with $z\approx 0.233$, ignoring their merger times.
In this case, most mergers including all range of eccentricites can be covered by all detectors.

%Most mergers have $e<0.1$ and their $\fpeak$ are in the detectable region of the space-borne detectors.
%There are three distinguishable groups of mergers with different $e$ as shown in the lower panel of Figure~\ref{fig:qlong}.
%Most of the mergers with circular orbits (dark blue curves) can be detected.
%They may also reach the detectable region of ground-based detectors \citep{Jani2020}.

%The low-$e$ group (light blue curves) has a fast decrease of the characteristic strain as $\fpeak$ increases. 
%They may only be detected by the space-borne detector. 
%The extremely eccentric group (the light green curves) cannot be detected. 
%Their evolution curves stop at a low strain.

\subsection{Estimation of IMBH-BH merger rate}
\label{sec:merge}
To estimate the IMBH-BH merger rate based on our models, the total stellar mass of Pop3 star clusters in the past must be obtained.
\cite{Tanikawa2022} provided a fitting formula (Equation 5) to the Pop3 star formation rate density from \cite{Skinner2020}.
The integration of the formula from 100 to 500~Myr (redshift from 20 to 10) indicates that the total stellar mass is approximately $3.2\times10^4 M_\odot \mathrm{Mpc}^{-3}$.
This estimation is an average value based on the numerical models of \cite{Skinner2020}.
\cite{Inayoshi2021} suggested that the upper li mit of Pop3 star formation rate inferred by re-ionization history is $2\times10^5 M_\odot \mathrm{Mpc}^{-3}$.
Assuming the average $\nmerge$ is 5 per 12~Gyr as shown in Figure~\ref{fig:mergebhavelong} for NFWden, the expected upper limit of IMBH-BH merger rate is $0.1 - 0.8~\mathrm{yr}^{-1} \mathrm{Gpc}^{-3}$ depending on the two star formation rate.
It is slightly larger than the estimated merger rate of $<0.01~\mathrm{yr}^{-1} \mathrm{Gpc}^{-3}$ from \cite{Tanikawa2021}, and comparable to the upper limit from \cite{LVK2021}. 
Considering the redshift dependence of the merger rate shown in Figure~\ref{fig:mflong}, the chance to detect GW events from the Pop3 star clusters at $z>6$ is expected to be considerably higher than the average rate.

\subsection{IMBH-star mergers}

%\com{AT}{Do IMBHs grow through IMBH-BH or IMBH-MS mergers? If IMBH-MS mergers happen frequently, it would be better to mention it. IMBH-MS (or IMBH-WD) mergers are also important for EM observations. It may be a future work.}
Generally, most of IMBHs in all types of environments grow their masses through IMBH-BH mergers. 
In contrast, part of them grow through mergers with main-sequence (MS), post main-sequence, and HeMS. 
%Such mergers can be observed as tidal disruption events \citep{Kashiyama2016}. 
%Note that they supposed direct collapse BHs with $\sim 10^5 M_\odot$ formed under Pop3 environments, which is more massive than IMBHs in our models by 2 or 3 orders magnitude. 
%Thus, such IMBH mergers may be too faint to observe.
Such mergers can trigger tidal disruption events launching relativistic jets \citep{Kashiyama2016}. We discuss about their detectability later in this section.

In the Pop3 star clusters with a top-heavy IMF, BHs are the major members of the cluster during the long-term evolution.
They occupy the center of the cluster.
Thus, IMBH-star mergers during the long-term evolution are unlikely to be formed, 
but they can occur at the early phase (within 10~Myr after the birth of stars).
We detected IMBH-star mergers in all short-term and long-term models, by selecting the mergers where $\mimbh>\mstar$.
For mergers with $\mimbh<\mstar$, we counted them as IMBH-VMS mergers in Table~\ref{tab:imbh-vms}. 
There is no IMBH-VMS in the long-term models.

\begin{figure}
	% To include a figure from a file named example.*
	% Allowable file formats are eps or ps if compiling using latex
	% or pdf, png, jpg if compiling using pdflatex
	\includegraphics[width=\columnwidth]{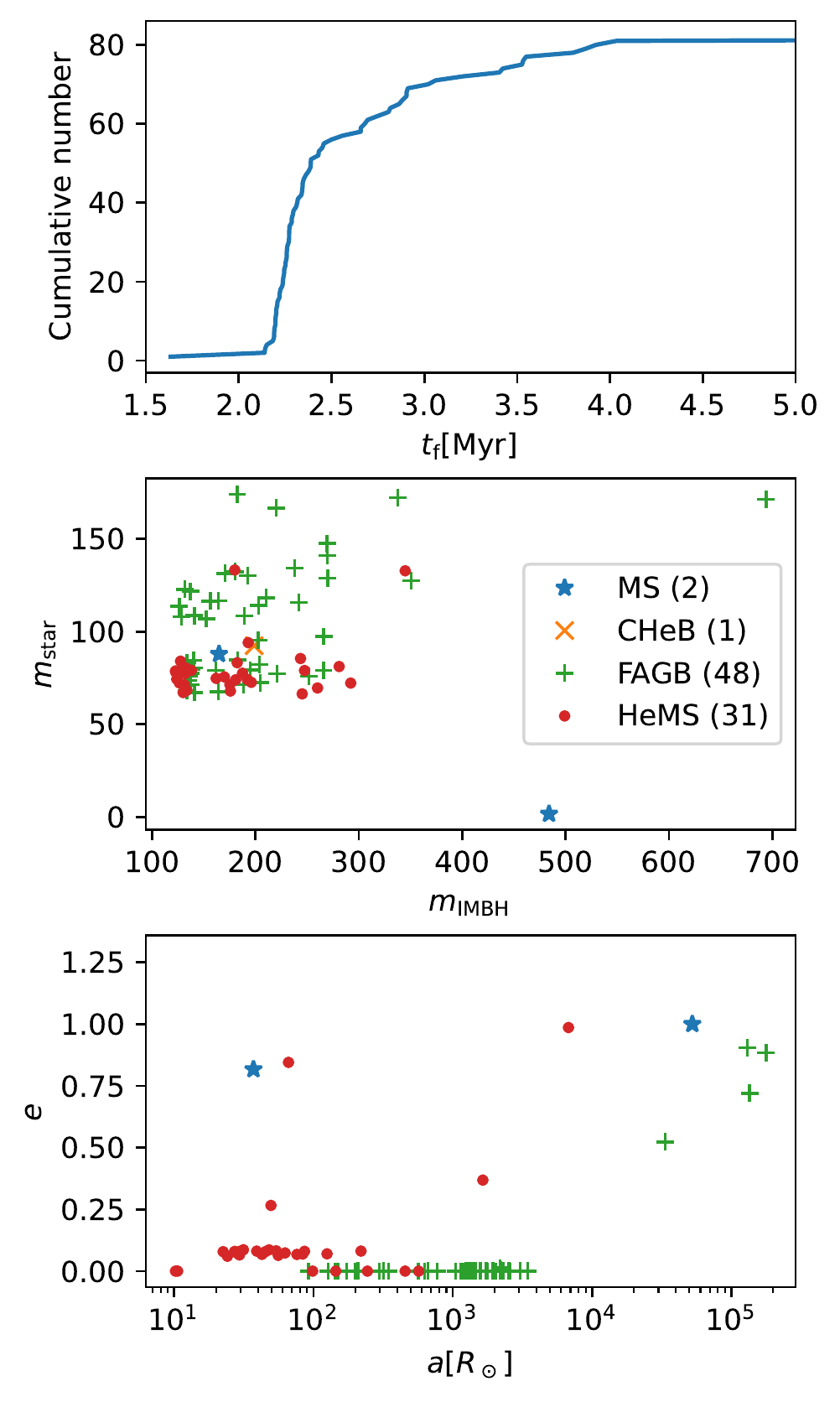}
    \caption{Properties of IMBH-star mergers. The upper panel: the cumulative distribution of $\tf$. The middle panel: Masses of stars ($\mstar$) v.s. masses of IMBHs ($\mimbh$). The lower panel: $a$ v.s. $e$. The meaning of symbols: MS: main sequence; CHeB: core Helium burning; FAGB: first asymptotic giant; HeMS: main-sequence naked Helium star.}
    \label{fig:imbhstar}
\end{figure}

Figure~\ref{fig:imbhstar} show the properties of IMBH-star mergers. 
Most of mergers occur between 2.0 to 2.5~Myr. 
The secondary stars can be MSs, core Helium burning stars (CHeBs), FAGBs, and HeMSs, where the last two are major ones.
The masses of FAGBs and of most HeMSs range from $60-180~M_\odot$ and $60-100~M_\odot$, respectively.
The primary IMBHs have masses below $400~M_\odot$, except two mergers of IMBH-MS and IMBH-FAGB.
Most IMBH-FAGB mergers have $a$ ranging from $100-3000~R_\odot$ and circular orbits, whereas most IMBH-HeMS mergers have smaller $a$ ranging from $10-600~R_\odot$ and nearly circular orbits.
A few high-eccentric orbits also exist.
Particularly, the two IMBH-MS mergers have $e>0.75$.
In the \textsc{bse} treatment, all IMBH-FAGB and IMBH-HeMS mergers are Thorne-Zytkow objects that IMBHs do not accrete material from companions. 

\cite{Kashiyama2016} suggested that prompt emission of relativistic jets launched by tidal disruption events will be observed in soft X-ray bands $\lesssim 10$ keV because of the high-redshift nature of Pop3 star clusters. They found that the peak isotropic luminosity of prompt emission is:
\begin{equation}
    L_{\mathrm{xray}} \sim 8 \times 10^{50} \mathrm{erg}~\mathrm{s}^{-1} \left( \frac{m_{\mathrm{IMBH}}}{10^5M_\odot} \right)^{-1/2} \left( \frac{m_{\mathrm{star}}}{40M_\odot} \right)^2 \left( \frac{R_{\mathrm{star}}}{3R_\odot} \right)^{-3/2},
\end{equation}
where they suppose that BHs are formed from supermassive stars with $\sim 10^5M_\odot$, and Pop3 MSs are disrupted. Figure \ref{fig:imbhstar} shows that $60-150M_\odot$ FAGBs and $60-100M_\odot$ HeMSs are disrupted by BHs with $100-300M_\odot$ in our Pop3 star clusters. The peak luminosity is $\sim 10^{48}$ erg~s$^{-1}$ in the FAGB case, as their stellar radii are $\sim 3000R_\odot$. 
In contrast, the peak luminosity can reach $\sim 10^{52}$ erg~s$^{-1}$ in the HeMS case because of their compactness. 
As the observed flux in the HeMS case is larger than $10^{-9}$ erg~s$^{-1}$~cm$^{-2}$ even at a redshift of 20, tidal disruption events of HeMSs in Pop3 star clustrers can be discovered by eROSITA with a limiting flux of $\sim 10^{-12}$ erg~s$^{-1}$~cm$^{-2}$ \citep{Merloni2012}.

\section{Discussion}
\label{sec:discussion}

The estimated merger rate is high as discussed in Section~\ref{sec:merge}.
However, our $N$-body models only cover a small parameter space of Pop3 star clusters. 
We discuss how the formation of VMS and the merger rate depend on a few uncertain factors

Firstly, the NFW profile is a simply approximation of real mini dark matter halos, which can grow up via mergers or can be disrupted by the tide from a massive host environment.
During a merger, two embedded star clusters may also merge.
Such an impact is difficult to estimate unless a self-consistent hydrodynamic model like that in \cite{Sakurai2017} is performed for 12~Gyr, which is particularly challenging due to the time-consuming computing.
The NFWden and the NFWstd dark matter halos have a density difference of approximately 8 times, whereas the $\nmerge$ only differ by one. The influence is not significant. 
If the halo grows up, the star cluster may be denser and generate more mergers.
If the halo is tidally disrupted or absorbed by the massive host environment, the star cluster might also dissolve. 
It is still possible that the final remnant of the system is an IMBH-BH binary that can merge today.

The formation of VMS is determined by the collision timescale described by Equation~\ref{eq:coll}, which depends on the initial central density of Pop3 star clusters.
For $W_0=6$ of King model, the formation of VMS does not occur, and no IMBH forms.
This density profile is close to the Plummer model, which is commonly used to describe the morphology of star clusters after dynamical virialization and before core collapse.
During the cluster formation stage, the morphology of the cluster is highly irregular with a high central density, as shown in the hydrodynamical simulation from \cite{Sakurai2017}. 
Thus, considering a more realistic condition, the chance to form a VMS in Pop3 star cluster is higher. 

We assume that all Pop3 stars form in the star clusters with $\Msc = 10^5 M_\odot$. 
This is probably not the case, a more realistic assumption is that the total mass of Pop3 star clusters follow a distribution peaking around $10^5 M_\odot$. For low-mass clusters, if the central density is sufficiently high ( $\rhorl>7\times10^5 M_\odot \mathrm{pc}^{-3}$), the multiple mergers of MS stars may occur and VMSs may form. 
\cite{Sakurai2017} show that all Pop3 star clusters with $\Msc$ from $5\times 10^4$ to $1.6\times 10^5 M_\odot$ satisfy this criterion, and there is no clear correlation between core densities of clusters and $\Msc$
In their model, the lowest $\Msc$ had the highest core density.
%Thus, we expect that the formation rate of VMSs in Pop3 star clusters 
Thus, we expect that the formation rate of VMSs in Pop3 star clusters with $\Msc>=1 \times 10^4 M_\odot$ to be high.
A part of low-mass Pop3 clusters may not form VMSs.
The threshold of $\Msc$ is unclear.
With a top-heavy IMF, few stars can form in a low-mass cluster.
Thus, we expect that the threshold to be around $10^3-10^4 M_\odot$.
Under the dark matter halo, all Pop3 star clusters can still survive and produce GW mergers.
A low-mass gas cloud may form a single VMS instead of a low-mass star cluster. 
Then, an IMBH can form but no GW merger occurs. 
Thus, our estimation of merger rate is an upper limit, considering such uncertainty.
%Under the strong gravitational potential from the dark matter halos, we expect that the low-mass clusters probably can also survive and produce mergers but the masses of IMBHs might be smaller.

The assumption of top-heavy IMF based on the theoretical models of Pop3 star formation may be updated to more realistic models in the future.
However, we expect that different IMFs weakly affect the merger rate. The models of \cite{Sakurai2017} assumed the canonical IMF \citep{Salpeter1955,Kroupa2001,Chabrier2003}, and also produced IMBH seeds. %Thus, the shape of IMF does not strongly affect the formation of IMBH. 
Mergers may occur with a top-heavy IMF, but the merger rate per BH would be lower compared to that of a canonical IMF because of a lower central density caused by the initial strong stellar winds from massive stars and the binary heating \citep{Wang2021}.
%Thus, we do not expect that a different IMF can significantly affect the merger rate.

The upper mass limit of IMF is 150~$M_\odot$ in our models, which was commonly assumed for Population I and II stars \citep{Kroupa2001}. 
For Pop3 stars, the upper mass can be considerably higher, even reaching 1000~$M_\odot$ if only a single star is allowed to form in a halo \citep[][]{Hirano2015,Hosokawa2016}. 
If the maximum mass is higher, the formation of IMBH becomes easier.

There is no primordial binary in our models.
Observations of the young star-forming region show that OB stars are mostly in multiple systems \citep{Sana2012}. 
It is unclear whether the formation of Pop3 stars follows the same tendency.
However, with primordial binaries, we expect that the mergers rate to be higher and the formation of VMSs to become easier. 
As shown in \cite{Wang2022}, the structure (density profile) evolution of star clusters weakly depends on the fraction of primordial binaries of OB stars.
With OB stars all in binaries initially, the dynamically driven mergers of BBHs increases.
Thus, we expect IMBH-BH mergers during the long-term evolution to also increase.

In our simulations, the mergers of BBHs are treated by the \textsc{bseemp} code using the orbital shrinking from \cite{Peters1964}. 
The GW kick is not considered.
When IMBH-BH merges, the high-velocity kick due to anisotropy of GW emission can eject the merger from the cluster.
The kick velocity depends on $q$ and the spins of BHs.
The mean kick velocity roughly linearly increases from 0 to 200~pc~Myr$^{-1}$in the range of $q$ from 0 to $0.2$, as shown in \cite{Morawski2018}, which is based on the formula from  \cite{Baker2008}.
$vesc$ is $50-70$~pc~Myr$^{-1}$ for our models with dark matter halos, as discussed in Section~\ref{sec:longterm}.
Thus, if the IMBH mass is not sufficiently large, an IMBH-BH merger may eject the IMBH from the cluster. 
This can reduce the IMBH-BH merger rate, whereas the BBH merger rate may not be considerably influenced.
If the dark matter halo is much denser than the NFWden halo, where $\vesc>200$~pc~Myr$^{-1}$, we expect that IMBHs can also be trapped and the merger rate is consistent with our model. 
The initial masses of IMBHs are not affected by the GW kick as they form via the stellar evolution of massive main sequence stars.

\section{Conclusions}
\label{sec:conclusion}

In this work, we conducted a series of $N$-body simulations for Pop3 star clusters embedded in static mini dark matter halos.
We investigated how VMSs and IMBHs form depending on the initial conditions of star clusters.
%and how dark matter halos affect the long-term evolution of star clusters.
We found that the dark matter halos can trap the Pop3 star clusters such that they can survive until today and continue to produce BBH and IMBH-BH mergers.
The low-$z$ mergers can be detected by the space-borne GW detectors, including LISA, TianQin, and Taiji, and most mergers can be detected by DECIGO, aLIGO, and Kagra.
The merger rate is high; thus, the GW events from Pop3 star clusters can be an important contribution to the future GW detection.
%We also calculate $\fpeak$ and the characteristic strain of mergers and find that they can 

We found that the central density is the major parameter that determines whether multiple mergers can occur in a short time ($\le 2$~Myr) to form a VMS.
The dynamically formed binaries via three-body channel are the major sources for the mergers.
For King model with $W_0=9$ and $12$, the collision rate is sufficiently high (Equation~\ref{eq:coll}) to drive multiple mergers within a short time to form a VMS before the stellar wind removes most of the mass (Figure~\ref{fig:mergetree}).
For $W_0=6$, the multiple mergers do not occur; thus, no VMS forms.
However, the high central density that can form a VMS as shown in \cite{Sakurai2017} is a more realistic condition at the early phase.
Other parameters, including the degree of primordial mass segregation, the lower mass boundary of IMF, and properties of dark matter halos, have almost no influence on the formation of VMS. 

The extremely metal-poor VMS eventually evolves to an IMBH with mass above $400~M_\odot$ via three possible modes: single stellar evolution, binary stellar evolution, and mass accretion in a BH-VMS system (Figures~\ref{fig:mmerge} and \ref{fig:mfshort}).
With a dark matter halo, the Pop3 star cluster can keep $\rh$ below $10$~pc for 12~Gyr (Figure~\ref{fig:rhrc}) because it significantly increases $\vesc$ of stars and suppresses the tidal evaporation of star clusters.
Thus, if the dark matter halo still survives today, a dark cluster might exist at the center and can produce BBH and IMBH-BH mergers, that might be observed by the GW detector.
Without dark matter halo, the cluster expands quickly in approximately 1~Gyr, the galactic tide can easily disrupt the cluster, similar to the fate of young star clusters with few thousands stars.
In this case, the cluster might leave an IMBH-BH system eventually. 

The estimation of the characteristic strain and $\fpeak$ indicates that a part of IMBH-BH and BBH mergers with low-$z$ in our models are above the sensitivity limit of LISA, TianQin and Taiji, and most mergers are covered by DECIGO, aLIGO and Karga in the high-frequency region (Figure~\ref{fig:obs}).
If we assume that mergers occur within 1~Gpc ($z\approx 0.233$), most events can be covered by all detectors.
We also found that the extremely high-eccentric mergers can form in Pop3 star clusters, where $1-e$ can reaches $10^{-8}$. 
They cannot be detected by LISA, TianQin, and Taiji, but can be interesting sources for other detectors. 
Mergers with PIBHs can also occur. 
The PIBHs can form via weak wind from extremely metal-poor stars, binary stellar evolution, and BBH mergers.

There are also mergers of IMBH-star binaries, where the stars can be MS, CHeB, FAGB, or HeMS, where the latter two are the major contributors. 
They can evolve to Thorne-Zytkow objects where IMBHs do not accrete material during the merging process.
The tidal disruption events of HeMSs can be discovered by eROSITA with a limiting flux of $\sim 10^{-12}$~erg~s$^{-1}$~cm$^{-2}$.

Based on the star formation rate of Pop3 stars from \cite{Skinner2020} and \cite{Inayoshi2021}, we estimate the upper limit of IMBH-BH merger rate is approximately $0.1 - 0.8~\mathrm{yr}^{-1} \mathrm{Gpc}^{-3}$.
The mergers can occur at any $z$ from 20 to 0, whereas a large fraction ($0.4-0.8$) occurs between 6 to 20 (Figure~\ref{fig:mflong}).
The PIBH-LBH merger rate is $0.01-0.15$~yr$^{-1}$~Gpc$^{-3}$, which is comparable to the merger rate from other channels.
Thus, the GW190521-like mergers are also possible from Pop3 star clusters.

Considering the GW kick, a part of IMBHs might be ejected after GW mergers, which might reduce the total merger rate. 
The disruption of dark matter halo can also reduce the merger rate.
If many Pop3 stars are in low-mass star clusters where $\Msc \ll 10^4 M_\odot$, the total number of IMBHs can also be lower.
Even considering this effect, the IMBH-BH mergers from Pop3 star clusters can still be an important contribution to the GW events that can be detected by the new generation of GW detectors. 

\section*{Acknowledgements}
L.W. thanks the support from the one-hundred-talent project of Sun Yat-sen University, the Fundamental Research Funds for the Central Universities, Sun Yat-sen University (22hytd09) and the National Natural Science Foundation of China through grant 12073090. 
L.W. also thanks the financial support from JSPS International Research Fellow (Graduate School of Science, The University of Tokyo).
This work is partly supported from the Grants-in-Aid for Scientific Research (17H06360, 19K03907).
We thank Yi-Ming Hu for providing the data of sensitivity curve of TianQin.

%%%%%%%%%%%%%%%%%%%%%%%%%%%%%%%%%%%%%%%%%%%%%%%%%%
\section*{Data Availability}
The simulations underlying this article were performed on the personal computing server of the first author.
The data were generated by the software \textsc{petar}, which is available in GitHub, at https://github.com/lwang-astro/PeTar.
The stellar evolution code \textsc{bseemp} is also available in GitHub https://github.com/atrtnkw/bseemp. 
The initial conditions of star cluster models are generated by the software \textsc{mcluster} \citep{Kuepper2011}, which is available in GitHub, at https://github.com/lwang-astro/mcluster.
The simulation data will be shared via private communication with a reasonable request. 

%%%%%%%%%%%%%%%%%%%% REFERENCES %%%%%%%%%%%%%%%%%%

% The best way to enter references is to use BibTeX:

\bibliographystyle{mnras}
\bibliography{ref} % if your bibtex file is called example.bib

% Alternatively you could enter them by hand, like this:
% This method is tedious and prone to error if you have lots of references
%\begin{thebibliography}{99}
%\bibitem[\protect\citeauthoryear{Author}{2012}]{Author2012}
%Author A.~N., 2013, Journal of Improbable Astronomy, 1, 1
%\bibitem[\protect\citeauthoryear{Others}{2013}]{Others2013}
%Others S., 2012, Journal of Interesting Stuff, 17, 198
%\end{thebibliography}

%%%%%%%%%%%%%%%%%%%%%%%%%%%%%%%%%%%%%%%%%%%%%%%%%%

%%%%%%%%%%%%%%%%% APPENDICES %%%%%%%%%%%%%%%%%%%%%

%\appendix
%
%\section{Some extra material}
%
%If you want to present additional material which would interrupt the flow of the main paper,
%it can be placed in an Appendix which appears after the list of references.

%%%%%%%%%%%%%%%%%%%%%%%%%%%%%%%%%%%%%%%%%%%%%%%%%%

% Don't change these lines
\bsp	% typesetting comment
\label{lastpage}
\end{document}